\documentclass[prl,10pt,twocolumn,tightenlines,superscriptaddress,numerical,notitlepage,showpacs,amsmath,amssymb,amsfonts,aps,longbibliography,floatfix]{revtex4-1}
\usepackage{graphicx}
\usepackage[colorlinks=True,linkcolor=red,citecolor=blue,urlcolor=blue]{hyperref}
\usepackage{bm}
\usepackage{bbm}
\usepackage{array}
\usepackage{braket}
\usepackage{cancel}
\usepackage{xcolor}
\usepackage{slashed}
\usepackage{comment}
\usepackage{combelow}
\usepackage{chngcntr}
\usepackage{nicefrac}
\usepackage{bookmark}
\usepackage[T1]{fontenc}
\usepackage[english]{babel}
\usepackage[normalem]{ulem}

\newcolumntype{C}{>{$}c<{$}}

\newcommand{\add}[1]{{#1}}

\newcommand{\beqn}{\begin{eqnarray}}
\newcommand{\eeqn}{\end{eqnarray}}
\newcommand{\beqs}{\begin{subequations}}
\newcommand{\eeqs}{\end{subequations}\\[-2mm]\noindent}
\newcommand{\eq}[1]{(\ref{#1})}

\newcommand{\bs}{\boldsymbol}
\newcommand{\avr}[1]{{\left\langle #1 \right\rangle}}

\newcommand{\lr}[1]{ \left( #1 \right) }

\definecolor{purple}{rgb}{0.8,0,0.6}
\definecolor{PURPLE}{rgb}{0.8,0,0.6}
\definecolor{orange}{rgb}{1,0.55,0}
\definecolor{limegreen}{rgb}{0.2,0.8,0.2}

\allowdisplaybreaks

\newcommand{\usefigs}{Figures_final3}
\newcommand{\usefigsSM}{Figures/response2}
\begin{document}
\date{\today}

\author{M. N. Chernodub}
\affiliation{Institut Denis Poisson UMR 7013, Universit\'e de Tours, 37200 Tours, France}
\affiliation{Department of Physics, West University of Timi\cb{s}oara, Bd. Vasile P\^arvan 4, Timi\cb{s}oara 300223, Romania}
\author{V. A. Goy}
\affiliation{Pacific Quantum Center, Far Eastern Federal University, 690922 Vladivostok, Russia}
\affiliation{\add{Institute of Automation and Control Processes, Far Eastern Branch, Russian Academy of Science, 5 Radio Str., Vladivostok 690041, Russia}}
\author{A. V. Molochkov}
\affiliation{Institut Denis Poisson UMR 7013, Universit\'e de Tours, 37200 Tours, France}
\affiliation{Pacific Quantum Center, Far Eastern Federal University, 690922 Vladivostok, Russia}
\affiliation{Beijing Institute of Mathematical Sciences and Applications, Tsinghua University, 101408, Beijing, China}
\author{D. V. Stepanov}
\affiliation{Pacific Quantum Center, Far Eastern Federal University, 690922 Vladivostok, Russia}
\affiliation{\add{Institute of Automation and Control Processes, Far Eastern Branch, Russian Academy of Science, 5 Radio Str., Vladivostok 690041, Russia}}
\author{A. S. Pochinok}
\affiliation{Pacific Quantum Center, Far Eastern Federal University, 690922 Vladivostok, Russia}

\ifdefined\compilesm\else
\title{Extreme Softening of QCD Phase Transition under Weak Acceleration:\\
First-Principles Monte Carlo Results for Gluon Plasma}

\begin{abstract}
We study the properties of gluon plasma subjected to a weak acceleration using first-principle numerical Monte Carlo simulations. We use the Luttinger (Tolman-Ehrenfest) correspondence between temperature gradient and gravitational field to impose acceleration in imaginary time formalism. Under acceleration, the system resides in global thermal equilibrium. Our results indicate that even the weakest acceleration up to $a \simeq 16$~MeV drastically softens the deconfinement phase transition, converting the first-order phase transition of a static system to a soft crossover for accelerating gluons. The accelerating environment can be relevant to the first moments of the early Universe and the initial glasma regime of relativistic heavy ion collisions. In particular, our results imply that the acceleration, if present, may also inhibit the detection of the thermodynamic phase transition from quark-gluon plasma to the hadronic phase.
\end{abstract}

\maketitle
\fi

\paragraph*{\bf Introduction.}
Quark-gluon plasma is a state of matter believed to have existed in the Early Universe up to a microsecond after the Big Bang~\cite{Rafelski2013}. In this phase, the fundamental constituents of matter, quarks and gluons, are not confined within hadrons. Relativistic collisions of heavy ions recreate this state in a laboratory setting, allowing experimental access to the fascinating properties of quark-gluon plasma~\cite{Pasechnik:2016wkt}.

When ions collide, they transfer their kinetic energy to a rapidly expanding plasma fireball, thus experiencing a rapid deceleration~\footnote{The deceleration is an acceleration with $a < 0$.}. The deceleration is caused by the strong longitudinal chromoelectric fields that mediate the interaction between the ions. The strength of these fields is of the order of $E \sim Q_s^2/g$, where \add{$Q_s {\simeq} 1$~GeV is the gluon saturation scale and} $g$ is the strong coupling constant. The typical deceleration achieved in a collision is argued to be of the order of the saturation scale, $|a| \simeq Q_s$~\cite{Kharzeev:2005iz}.

A uniformly accelerated system possesses the Rindler event horizon~\cite{Rindler:1966zz} beyond which the events do not influence the accelerating particles~\cite{Lee1986}. The presence of the event horizon fosters the Unruh effect, where an observer, uniformly accelerated in a zero-temperature Minkowski vacuum, detects a thermal bath of particles with the Unruh temperature:~\cite{Unruh:1976db}
\begin{align}
 T_U = \frac{|a|}{2\pi}\,.
\label{eq_T_U}
\end{align}
The Rindler horizon of an accelerated system has a profound similarity with the event horizon of a black hole, which also separates causally disconnected regions of spacetime.~\cite{Gibbons:1976pt, Gibbons:1976es} For black holes, the presence of the event horizon leads to a very intriguing quantum effect: black holes evaporate via the Hawking production of particle pairs, where one of the particles falls to the black hole while the other one escapes to the spatial infinity.~\cite{Hawking1974, Hawking1975} The particle radiation, which can be interpreted as a tunneling process through the event horizon~\cite{Parikh:1999mf}, is perceived as thermal radiation with the Hawking temperature $T_H = \kappa/(2\pi)$, where $\kappa=1/(4M)$ is the acceleration due to gravity at the event horizon of a black hole with mass $M$. The similarity of the Hawking and Unruh temperatures~\eqref{eq_T_U} suggests that the thermal character of both phenomena originates from the presence of the appropriate event horizons.

In application to colliding ions, deceleration was suggested to lead to a rapid thermalization of the gluon matter through the Unruh effect, which was argued to produce a final thermal gluon state via quantum tunneling through the emerging event horizon. The effect should appear at the short time scale of $t \simeq 2\pi/Q_s \simeq 1$~fm with the resulting heat bath temperature~\eqref{eq_T_U} reaching a typical QCD scale of $T = |a|/(2\pi) \sim 200$~MeV.~\cite{Kharzeev:2005iz} The rapid thermalization leads to the subsequent emergence of quark-gluon plasma, which achieves thermal equilibrium at a later stage.

Besides the thermalization process, the acceleration of interacting particle systems can cause changes in their phase structure. The restoration of the chiral symmetry for interacting fermions has been considered in Ref.~\cite{Ohsaku:2004rv} within the Nambu--Jona-Lasinio model~\cite{Nambu:1961tp, Nambu:1961fr}. This effect seems to be a natural consequence of the observation that the acceleration is associated with the Unruh temperature~\eqref{eq_T_U}, while thermal effects usually lead to the restoration of a spontaneously broken symmetry.~\cite{Dolan:1973qd} The symmetry restoration due to acceleration has been questioned in Ref.~\cite{Unruh:1983ac}, which concluded that the acceleration of a vacuum itself cannot cause phase transitions. While we will not delve into this question, we mention that one should distinguish two physical scenarios of acceleration: (i) a vacuum of interacting particles viewed from the point of the acceleration observer and (ii) a physically accelerated object from the points of view of the observer that accelerates together with the object. While Ref.~\cite{Unruh:1983ac} deals with the former question, here we consider an accelerated thermal state of hot gluons at $T \geqslant T_U$, which corresponds to the thermally equilibrated system of accelerated particles. For an equilibrium system, the region with $T < T_U$ is usually considered to be forbidden~\cite{Becattini:2017ljh, Prokhorov:2019cik, Prokhorov:2019sss, Palermo:2021hlf} (see, however, a recent discussion in Refs.~\cite{Prokhorov:2023dfg, Prokhorov:2024hjb}). Restoration of a spontaneously broken symmetry by acceleration has also been addressed in various physical scenarios~\cite{Ebert:2006bh, Castorina:2012yg, Takeuchi:2015nga, Dobado:2017xxb, Casado-Turrion:2019gbg, Kou:2024dml}. A very recent critical assessment of the restoration of broken symmetry due to acceleration in an interacting field theory can be found in Ref.~\cite{Salluce:2024jlj}.

In our work, we study the non-perturbative properties of gluon plasma subjected to weak acceleration using first-principle numerical Monte Carlo simulations. Throughout this article, we use the units $c = \hbar = 1$.

\vskip 1mm
\paragraph*{\bf Global thermal equilibrium under acceleration.}

Under a uniform acceleration, a generic particle system resides in global thermal equilibrium characterized by inhomogeneous temperature $T(x)$. In a classical approach, it is convenient to describe the corresponding physical environment by the inverse temperature four-vector $\beta^\mu(x) \equiv u^\mu(x) / T(x)$, which is associated with the local fluid velocity $u^\mu = u^\mu(x)$.  

In thermal equilibrium, the inverse temperature $\beta^\mu$ satisfies the Killing equation, $\partial_\mu {\beta}_\nu + \partial_\nu {\beta}_\mu = 0$~\cite{Cercignani:2002, Becattini:2012tc}. For a thermalized system that accelerates uniformly in the direction $z$, one gets the appropriate solution of this equation: $\beta^\mu(x) \partial_\mu = (1/T_0) [(1 + a_0 z) \partial_t + a_0 t \partial_z]$. Then, the local temperature $T(x)$, the local fluid velocity $u^\mu(x)$, and the local proper acceleration $a^\mu(x) \equiv u^\nu \partial_\nu u^\mu$ of the accelerated particle fluid are respectively (see, {\it e.g.}, \cite{Ambrus:2023smm}):
\begin{subequations}
\begin{align}
 T(t,z) & = \frac{T_0}{\sqrt{(1 + a_0 z)^2 - (a_0 t)^2}}\,,
\label{eq_T_z} \\ 
u^\mu(t,z) \partial_\mu & = \frac{T(x)}{T_0} 
\bigl[(1 + a_0 z) \partial_t + a_0 t \partial_z\bigr]\,,
	\label{eq_u}\\
    a^\mu(t,z) \partial_\mu & = a_0 \frac{T^2(x)}{T_0^2} [a_0 t \partial_t + (1 + a_0 z) \partial_z]\,,
\label{eq_a}
\end{align}
\label{eq_Tua}
\end{subequations}
~\hskip -3mm 
where $T_0 = T(t = z = 0)$ and $a_0 = a(t = z = 0)$ are interpreted as the reference quantities taken at a $z = 0$ plane at time $t = 0$. The proper acceleration, $\alpha^\mu = a^\mu(x)/T(x)$, has a constant magnitude, $\alpha^\mu \alpha_\mu = - a_0^2/T^2_0$. 

For definiteness, we take $a_0 > 0$. Then, an accelerating particle follows a hyperbolic trajectory in spacetime with the entire worldline confined within the right Rindler wedge, $z > |t| - 1/a_0$. The boundary of the Rindler wedge corresponds to the Rindler horizon,
\begin{align}
    z_{\rm R}(t) = |t| - 1/a_0 \,,
\label{eq_Rindler_horizon}
\end{align}
which sets the physical causal boundary of the system. Any events that appear beyond the Rindler horizon, at $z < z_{\rm R}(t)$, cannot affect the particle motion because the light signals from those events will never reach the particle subjected to a constant acceleration. Consequently, the thermodynamics of a uniformly accelerating particle system is defined only within the right Rindler wedge. At the Rindler horizon~\eqref{eq_Rindler_horizon}, all quantities~\eqref{eq_Tua} diverge.

It is convenient to rewrite Eqs.~\eqref{eq_Tua} at $t = 0$, which give the local temperature and acceleration, $a^\mu(z) = a(z) \delta^{\mu,3}$,
\begin{align}
    T(z) = \frac{T_0}{1 + a_0 z}\,, 
    \qquad
    a(z) = \frac{a_0}{1 + a_0 z}\,,
\label{eq_Ta}
\end{align}
for a locally static fluid $u^\mu = \delta^{\mu,0}$. Equations~\eqref{eq_Ta} show that acceleration is tied to a temperature gradient,
\begin{align}
    a(z) = - \frac{1}{T(z)} \frac{\partial T(z)}{\partial z}\,,
    \label{eq_TE}
\end{align} 
in accordance with the Tolman-Ehrenfest~\cite{Tolman:1930zza, Tolman:1930ona} and Luttinger~\cite{Luttinger:1964zz}, relations. The Rindler horizon~\eqref{eq_Rindler_horizon} then becomes simply $z_{\rm R} = - 1/a_0$.

We study the effect of a weak acceleration on the gluon plasma in the first-principle approach within the lattice gauge theory. We implement the $T = T(z)$ temperature profile~\eqref{eq_Ta} at a fixed $z$-independent spatial volume, noticing that our $t = 0$ results can be extended to any $t$ by replacing the $a(z) \to a(t,z)$ in the calculated expectation values~\cite{Becattini:2019poj, Becattini:2020qol, Selch:2023pap}. The path-integral formalism justifying our setup has been carefully elaborated in Ref.~\cite{Selch:2023pap}. 

\vskip 1mm
\paragraph*{\bf No sign problem for an accelerating system.}

In sharp contrast to vorticity and finite baryonic chemical potential, the accelerating environment can be directly implemented in the imaginary time formalism without the need to work in a complex plane and subsequently perform an analytical continuation to the real values of acceleration. Indeed, the acceleration is the rate at which the velocity ${\boldsymbol{v}} = \partial {\boldsymbol{x}}/\partial t$ of a body changes over time, ${\boldsymbol{a}} = \partial {\boldsymbol{v}}/\partial t \equiv \partial^2 {\boldsymbol{x}}/\partial t^2$. The Wick rotation from the real to imaginary time, $t \to - i \tau$, amounts to simply flipping the sign of the acceleration vector, $\boldsymbol{a} \to \boldsymbol{a}_{\rm E} = - \boldsymbol{a}$. Thus, we set the temperature gradient~\eqref{eq_Ta} in the imaginary-time formalism as $T(z) = T_0/(1 - a_E z)$. Hereafter, we denote $a = - a_E \equiv a_0$ to simplify our notations.

We also notice that the acceleration in the imaginary time formalism can be implemented,  respecting the Kubo–Martin–Schwinger (KMS) condition~\cite{Kubo1957, Martin1959}, at least in two other ways: (i) as a motion along the imaginary circle in the Euclidean spacetime and (ii) as a double-periodicity condition along imaginary time direction in the Euclidean-Rindler spacetime~\cite{Ambrus:2023smm}. Here, we use a third method of utilizing Tolman-Ehrenfest (Luttinger) correspondence~\cite{Tolman:1930ona, Tolman:1930zza, Luttinger:1964zz} set by the inhomogeneous temperature profile~\eqref{eq_Ta}.

\vskip 1mm
\paragraph*{\bf Lattice setup.}

We consider SU(3) gauge theory with the anisotropic Wilson action~\cite{Karsch:1982ve}:
\begin{align}
    S = & \sum_x \sum_{i>j=1}^3 {\beta}_{\sigma}(x_3) \left( 1 - {\mathcal P}_{x,ij} \right) 
    \nonumber \\ 
        & + \sum_x \sum_{i=1}^3 {\beta}_{\tau}(x_3) \left( 1 - {\mathcal P}_{x,4i} \right)\,,
    \label{eq_S_Karsch}
\end{align}
where $x = \{x_1, x_2; x_3 \equiv z; x_4 \equiv \tau\}$ is the Euclidean coordinate of the lattice sites, ${\mathcal P}_P = \frac{1}{3}\mathrm{Re\,Tr}\, U_P$, with $U_{x,\mu\nu} = U_{x,\mu} U_{x+{\hat\mu},\nu} U^\dagger_{x+{\hat\nu},\mu} U^\dagger_{x,\nu}$, is the lattice action on an elementary plaquette $P \equiv P_{n,\mu\nu} = \{n,\mu\nu\}$, and $\mu,\nu = 1, \dots, 4$ are Euclidean directions. We consider the lattices $N_\tau {\times} N_x {\times} N_y {\times} N_z$ with the size $N_\tau {=} 8$ along the imaginary time $\tau$, the transverse spatial size, $N_s \equiv N_x = N_y = \add{84}$ and the longitudinal sizes $N_z = 104,\,126,\,148,\,170$ along the acceleration direction~$z \equiv x_3$.

The Wilson action~\eqref{eq_S_Karsch} with two couplings, ${\beta}_{\sigma}$ and ${\beta}_{\tau}$, allows us to fix separately the spatial, ${\mathfrak a}_\sigma = {\mathfrak a}_\sigma({\beta}_\sigma, {\beta}_\tau)$, and temporal, ${\mathfrak a}_\tau =  {\mathfrak a}_\tau({\beta}_\sigma, {\beta}_\tau)$, lattice spacings that stand for the physical length of the lattice links in the appropriate directions.~\footnote{To avoid confusion, we use the Fraktur font $\mathfrak a$ for a lattice spacing and the Italic font $a$ for an acceleration.} We use an improved procedure following Ref.~\cite{Karsch:1982ve} to simulate the theory on the asymmetric lattices\add{~\cite{Athenodorou:2020ani}}. The technical details of our simulations are given in the Supplemental Material~\cite{SM}, which includes Refs.~\cite{Shirogane:2016zbf,Yang:2024tfc}.

We impose periodic boundary conditions along $x,\, y,\, \tau$ directions and keep open boundaries along the acceleration direction $z$. The physical length of the imaginary time circle, $L_\tau \equiv N_\tau {\mathfrak a}_\tau(z) = 1/T(z)$ is chosen to match the temperature profile~\eqref{eq_Ta} corresponding to a constant proper acceleration. In the $z$ direction, temperature varies only on a central segment that occupies half of the volume. We keep two $\delta L_z = {\mathfrak a}_\sigma N_z/4 \sim (1 \dots 4)\,{\rm fm}$ thick slices at both sites of the central segment to mitigate the effects of the open boundary conditions along the $z$ direction. It is important to stress that we made the lengths of the transverse spatial directions $L_i = N_i {\mathfrak a}_\sigma(z)$ with $i = x,y$ are made $z$-independent by imposing a particular $z$ dependence for the ${\beta}_{\sigma}$, and ${\beta}_{\tau}$ couplings in the lattice action~\eqref{eq_S_Karsch} (the details are given in Ref.~\cite{SM}).

In order to probe the phase structure of the model, we calculate the order parameter of confinement, the Polyakov loop, $L(z)$, and its susceptibility $\chi(z)$,
\begin{align}
    & L(z)  = \avr{|L_z|}\,, \qquad L_z = \frac{1}{N_s^2}\sum_{x,y=0}^{N_s - 1}\frac{1}{3} {\mathrm{Tr}\,}  \prod_{\tau = 0}^{N_\tau - 1} U_{{\bs x},\tau}\,, 
    \label{eq_L}\\ 
    & \chi(z)  = \avr{|L_z|^2} - \avr{|L_z|}^2 \,,
    \label{eq_chi}
\end{align}
where ${\bs x} = (x,y,z)$ is the spatial coordinate. The notation $\avr{{\cal O}}$ represents a statistical averaging of the quantity $\cal O$ over the whole ensemble of the field configurations. For each lattice size and acceleration, the central temperature at $z=0$ is chosen to be equal, within good numerical accuracy, to the critical temperature of the non-accelerating system, $T_0 \equiv T(z = 0)  = T_c(a=0)$.

\vskip 1mm
\paragraph*{\bf Results.}

Figure~\ref{fig_Polyakov_loop} shows the Polyakov loop in an accelerating gluon fluid. The results agree qualitatively with our expectations given the explicit form of the temperature profile~\eqref{eq_Ta}: the $a > 0$ acceleration enhances the deconfining phase at $z < 0$ and drives the system deeper into the confining phase at $z > 0$. 
\begin{figure}
    \centering
    \includegraphics[width=1.0\linewidth]{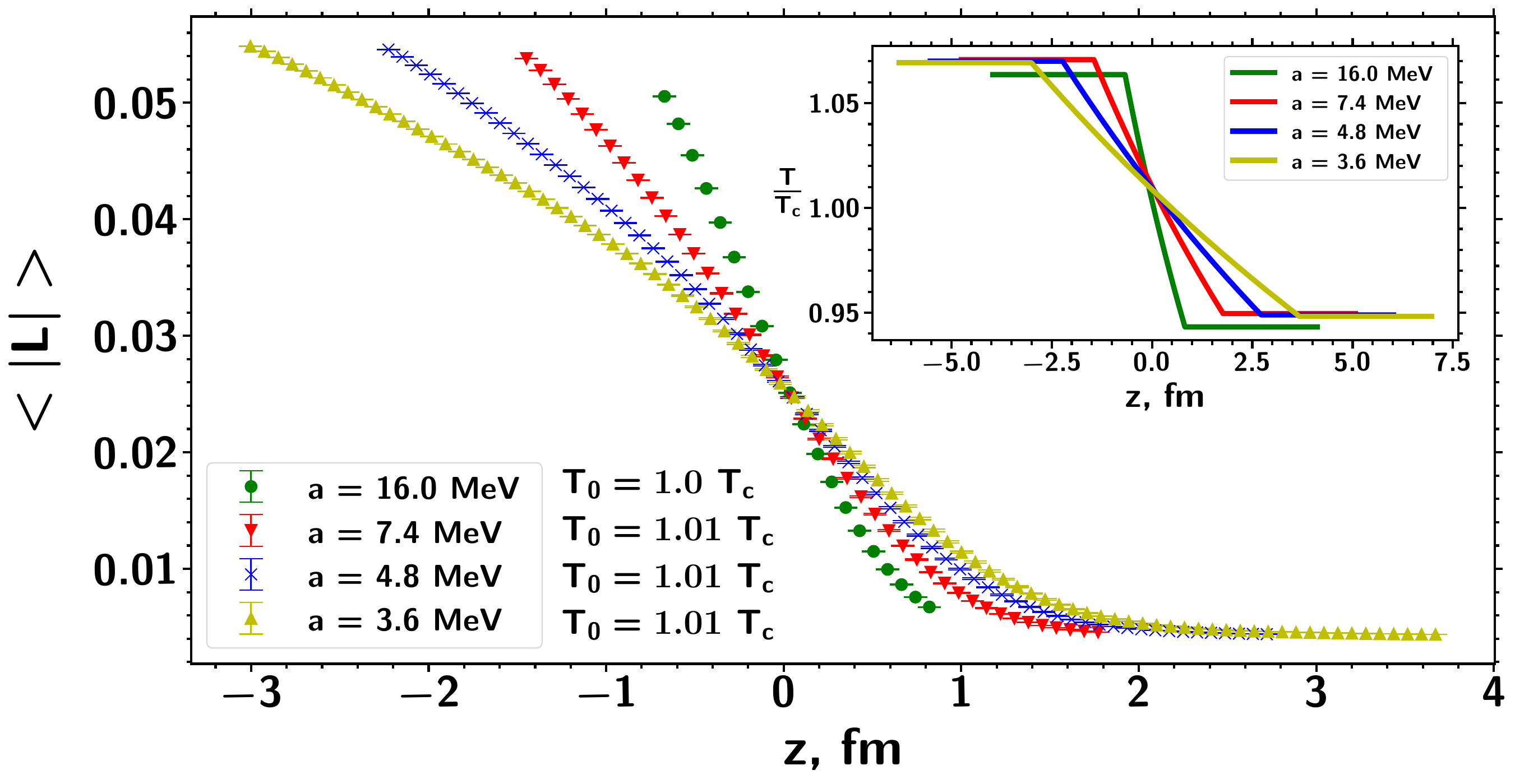}
    \caption{The local expectation value of the Polyakov loop, $L = L(z)$, at various accelerations $a \equiv a_0 = - a_{\rm E} > 0$ along the $z$ direction. Shown is a central region, where the temperature is inhomogeneous ({\it cf.} the inset). The temperature in the central plane, $z=0$, is chosen to be close to the critical deconfinement temperature $T_c$ without acceleration, $a=0$. The confining (deconfining) regions are located at the right, $z \gtrsim 0$ (left, $z \lesssim 0$) semi-volumes.}
    \label{fig_Polyakov_loop}
\end{figure}

To estimate the quantitative effect of acceleration on gluon medium, we compute the excess of the local free energy of an accelerating heavy quark at the position~$z$ with respect to the energy of a static ($a=0$) heavy quark that resides at the same temperature $T_{\rm bulk} = T(z, a)$ but now in the whole space:
\begin{align}
    \Delta F(z) = - T_{\rm bulk} \ln \frac{L_{\rm acc}(z,a)}{L_{\rm bulk}(T_{\rm bulk})}{\biggl|}_{T_{\rm bulk} = T(z,a)}\,,
    \label{eq_Delta_F}
\end{align}
with \add{$L_{\rm bulk} = \avr{L(z)}_z$ computed for the vanishing acceleration $a=0$ and temperature $T_{\rm bulk}$}.

The free energy, shown in Fig.~\ref{fig_Free_energy}, has a clear tendency to increase at the deconfining side ($z \lesssim 0$) of the accelerating gluon fluid. In other words, the acceleration tends to drive the deconfining plasma towards the confinement region. A reciprocal effect is observed in the deconfining side ($z \gtrsim 0$), where the acceleration makes the free energy lower, thus driving the system toward deconfinement. 
\begin{figure}
    \centering
    \includegraphics[width=1.0\linewidth]{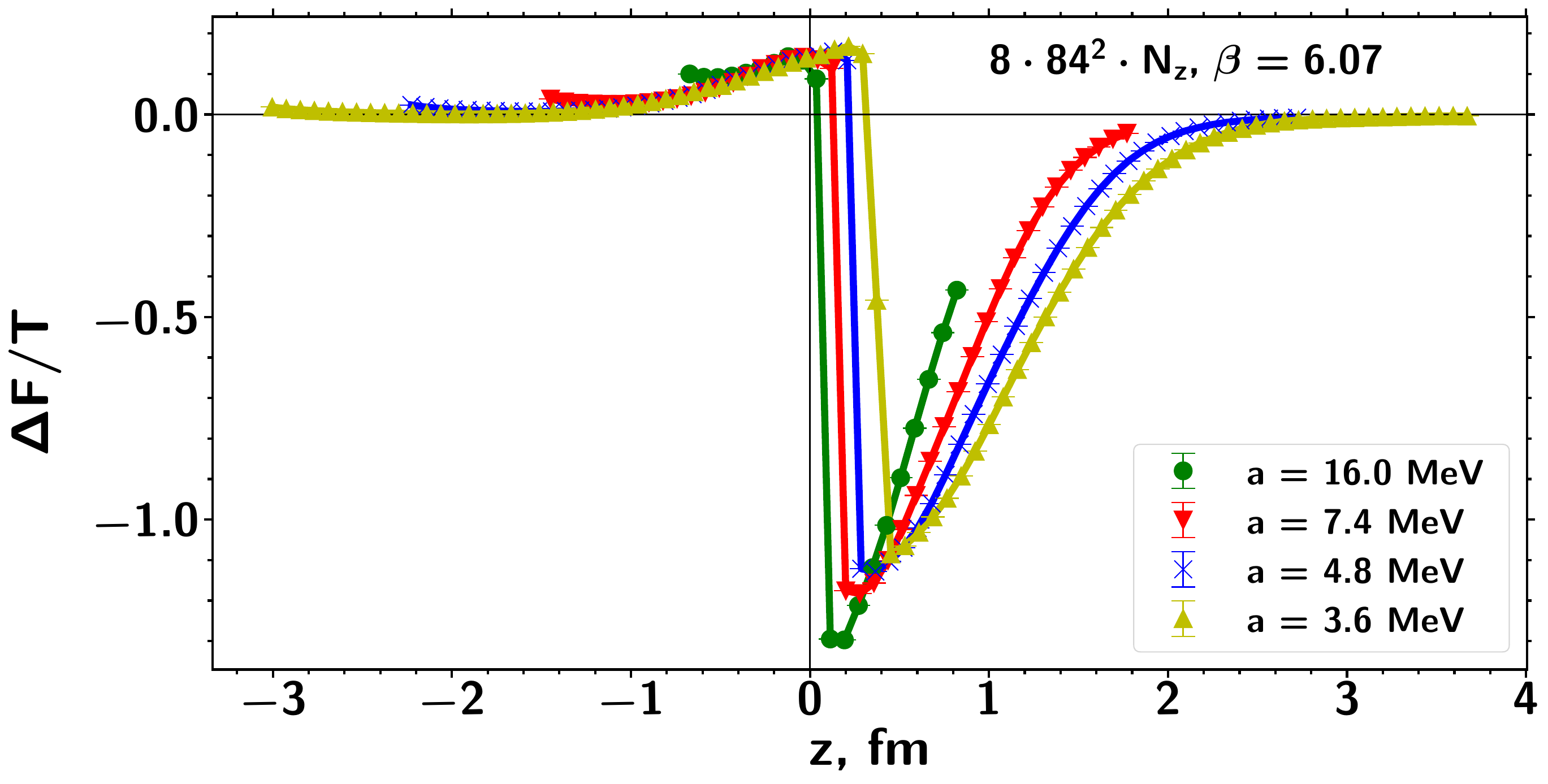}
    \caption{The effect of the acceleration on the free energy of a free heavy quark~\eqref{eq_Delta_F} with the notations of Fig.~\ref{fig_Polyakov_loop}.}
    \label{fig_Free_energy}
\end{figure}

In Figure~\ref{fig_Polyakov_susceptibility}, we show the susceptibility of the Polyakov loop at various accelerations. To compare this quantity at the non-accelerating case, we introduce the matching susceptibility, which is measured for the homogeneous, $z$-independent gluon matter at temperature $T_{\rm bulk}$ fixed to match $T = T(z, a)$ for chosen $z$ and $a$ of the accelerating gluon matter. For the matching acceleration, we took the moderate value $a \simeq 5$~MeV. Figure~\ref{fig_Polyakov_susceptibility} shows that a nonzero acceleration makes the phase susceptibility curve (i) wider and (ii) lower, as compared to the matching $a=0$ case, where the bulk variation of temperature is only taken into account. Both properties coherently imply that the accelerating gluon fluid experiences a broad crossover transition to the deconfining phase instead of the first-order phase transition at zero acceleration.

\begin{figure}
    \centering
    \includegraphics[width=1.0\linewidth]{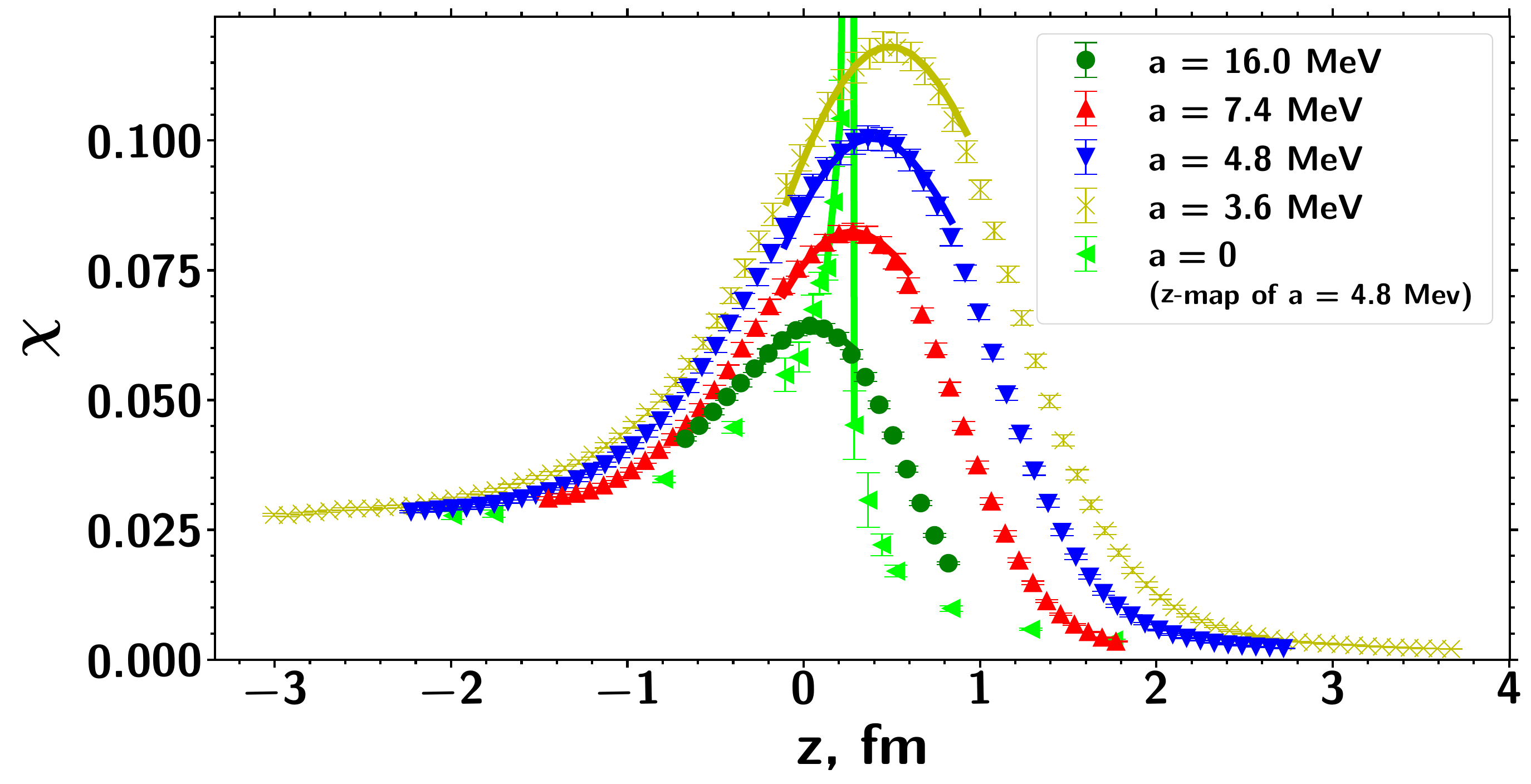}
    \caption{Susceptibility of the Polyakov loop~\eqref{eq_chi} for the accelerating gluon medium (in the notations of Fig.~\ref{fig_Polyakov_loop}). The matching $a=0$ curve shows the susceptibility $\chi = \chi(z)$ found at a globally homogeneous temperature $T_{\rm bulk}$ that matches the local temperature $T = T(z, a \simeq 5\, {\rm MeV})$ for each~$z$.}
    \label{fig_Polyakov_susceptibility}
\end{figure}

\add{To quantify the effect of acceleration on hot gluon matter, we fit the Polyakov loop susceptibility with a Gaussian function at each acceleration $a$. We identify the critical temperature with the position of the maximal value of the susceptibility and calculate the corresponding width of the deconfining transition. The fits were obtained by selecting a symmetric region around the maximum selected to saturate the standard deviation $\chi^2/{\rm d.o.f} < 1$. In Fig.~\ref{fig_Polyakov_susceptibility}, the solid lines represent the best fits for the lattice with the transverse extension $N_x = N_y = 84$. The phase diagram of the accelerating gluon matter is shown in Fig.~\ref{fig_Phase_Diagram} in the thermodynamic limit (see~\cite{SM} for details).}

\begin{figure}
    \centering
    \includegraphics[width=1.0\linewidth]{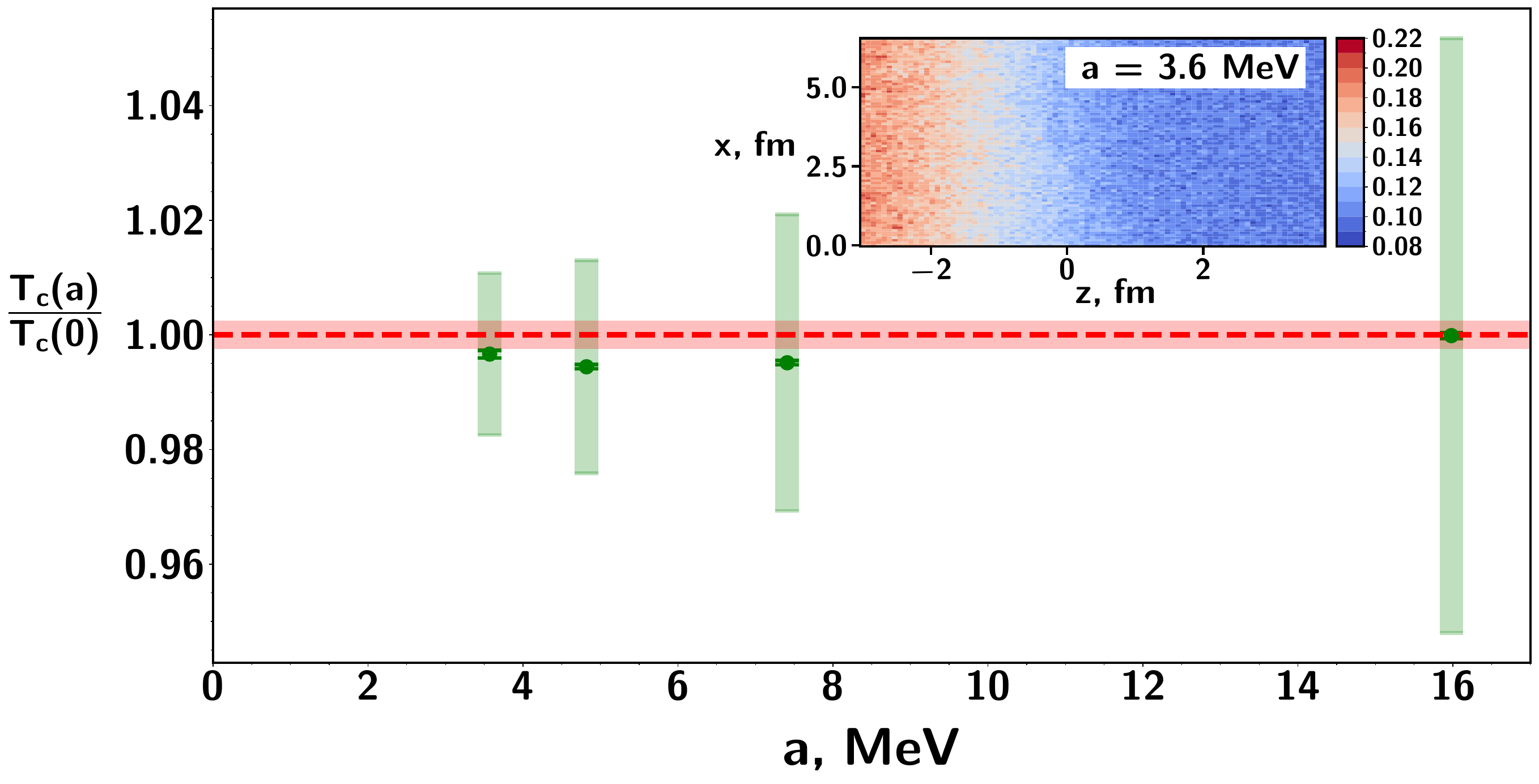}
    \caption{Phase diagram of the hot gluon matter under acceleration in the $(a, T)$ plane. \add{The data are presented for the infinite volume limit $N_{x,y}\rightarrow\infty$.} 
    The widths of the crossovers are shown by the semi-transparent bands. The inset depicts the density plot of the Polyakov line in the $(x,z)$ plane at a typical gluon configuration, averaged over the $y$ direction.}
    \label{fig_Phase_Diagram}
\end{figure}

Despite the physical value of acceleration $a$ being much smaller than the QCD mass scale, $a \ll \Lambda_{\rm QCD} \simeq 100\, {\rm MeV}$, the effect of the acceleration on the phase diagram is unexpectedly substantial. Our results show that even the weakest studied acceleration, $a \simeq 4\, {\rm MeV}$, converts the weak first-order thermal phase transition in SU(3) Yang-Mills theory into a smooth crossover. As the acceleration increases towards the biggest studied values $a \simeq 16\, {\rm MeV}$, the width of the crossover becomes larger. \add{The scaling towards infinite volume, presented in Ref.~\cite{SM}, reaffirms the crossover nature of the transition.}

The acceleration of gluon plasma does not affect the position of the (pseudo)critical temperature of the deconfinement crossover transition, which remains approximately equal, within a few percent accuracy, to the critical temperature of the phase transition of the non-accelerating system, $T^{\rm crossover}(a > 0) \simeq T^{\rm 1st\, order}_c(a=0)$.

\vskip 1mm
\paragraph*{\bf Conclusions.}
In our article, we studied the effects of acceleration on the phase diagram of gluon matter using first-principle Monte Carlo simulations. The thermal equilibrium of the gluon matter is ensured by the particular form of the spatial inhomogeneity~\eqref{eq_Ta} of local temperature dictated by the Killing equation. The temperature inhomogeneity satisfies the Luttinger (Tolman-Ehrenfest) correspondence~\eqref{eq_TE} between temperature gradient and gravitational field. On a practical side, we notice that the imposing of acceleration in the imaginary time formalism is free from the sign problem. Therefore, our approach allows us to calculate properties of accelerating systems directly in lattice Monte Carlo simulations.

We show that even the weakest acceleration of the order of $a \simeq 4\,{\rm MeV}$ drastically softens the deconfinement phase transition, converting the first-order phase transition of a static hot gluon system to a very soft and broad crossover without noticeably (with an accuracy within a few percent) shifting the position of the corresponding pseudocritical temperature. 
\add{Our results suggest the existence of a critical acceleration, $a = a_c$, at which the weak first-order phase transition of hot SU(3) gluon matter turns into a crossover. The position of the critical point can be probed with the latent heat measurements following, for example, Ref.~\cite{Shirogane:2016zbf}.}

The softening effect on the phase transition may have some consequences in the physical environments where (quark) gluon plasma experiences acceleration. One immediate application could be the early Universe. The softening could also hinder the search for the critical line in experiments in relativistic heavy-ion collisions. 

\vspace{2ex}
\begin{acknowledgments}
The work of MNC was funded by the EU’s NextGenerationEU instrument through the National Recovery and Resilience Plan of Romania - Pillar III-C9-I8, managed by the Ministry of Research, Innovation and Digitization, within the project entitled ``Facets of Rotating Quark-Gluon Plasma'' (FORQ), contract no.~760079/23.05.2023 code CF 103/15.11.2022. VAG and AVM have been supported by RSF (Project No. 23-12-00072)~\cite{RSFurl}.
AVM thanks the Institut Dennis Poisson (Tours, France) for the kind hospitality and acknowledges the support of Le Studium (Orl\'eans, France) research professorship. DVS and ASP were supported by Grant No. FZNS-2024-0002 of the Ministry of Science and Higher Education of Russia. The numerical simulations were performed at the computing cluster of Far Eastern Federal University and the equipment of Shared Resource Center "Far Eastern Computing Resource" IACP FEB RAS~\cite{IACPurl}.
\end{acknowledgments}

\bibliography{acceleration}

\clearpage
\newpage
\setcounter{equation}{0}
\setcounter{figure}{0}
\renewcommand{\theequation}{SM.\arabic{equation}}
\renewcommand{\thefigure}{SM\arabic{figure}}
\setcounter{page}{1}
\renewcommand{\thepage}{SM.\arabic{page}}

\ifdefined\compilesm
\title{
{\bf\Large{Supplemental Material for}}\\[3mm]
``Extreme Softening of QCD Phase Transition under Weak Acceleration:\\
First-Principles Monte Carlo Results for Gluon Plasma''
}

\maketitle
\else
\section{Supplemental Material}
\fi

\subsection{Numerical simulations}

The Wilson action~\eqref{eq_S_Karsch} with two couplings ${\beta}_{\sigma}$ and ${\beta}_{\tau}$ allows us to fix separately the spacial, ${\mathfrak a}_\sigma = {\mathfrak a}_\sigma({\beta}_\sigma, {\beta}_\tau)$, and temporal, ${\mathfrak a}_\tau =  {\mathfrak a}_\tau({\beta}_\sigma, {\beta}_\tau)$ lattice spacings. The quantity 
\begin{align}
    \xi = \frac{{\mathfrak a}_\sigma}{{\mathfrak a}_\tau}\,,
    \label{eq_xi}
\end{align}
gives us the anisotropy of the lattice spacings.

In the continuum limit, where both couplings ${\beta}_{\sigma}$ and ${\beta}_{\tau}$ are large, the lattice spacings vanish, ${\mathfrak a}_\sigma\to 0$ and ${\mathfrak a}_\tau\to 0$. Consequently, the lattice plaquette ${\mathcal P}_{x,\mu\nu}$ becomes proportional to the square of the continuum field-strength tensor $F_{\mu\nu}$ times a volume element, ${\mathfrak a}_\sigma^3 {\mathfrak a}_\tau$, and the lattice Wilson action~\eqref{eq_S_Karsch} reduces to the continuum action of the Yang-Mills theory~\cite{Karsch:1982ve}.

We work with asymmetric lattices, $N_\tau < N_\sigma$, corresponding to the case of finite temperature. The local temperature is given by an inverse of the lattice length $L_\tau = N_\tau {\mathfrak a}$ in the imaginary time direction:
\begin{align}
    T(z) = \frac{1}{N_\tau {\mathfrak a}_\tau(z)} \,.
    \label{eq_T}
\end{align}

We aim to introduce the temperature gradient along the longitudinal spatial direction $z \equiv x_3$:
\begin{align}
    T(z) = \frac{T_0}{1 + a_0 (z-z_0)}\,, 
    \label{eq_T_x3}
\end{align}
where $T_0 \equiv T(z{=}z_0)$ is the temperature at the $z = z_0$ plane and $a_0 > 0$ is the parameter of the temperature inhomogeneity which corresponds, according to the Tolman-Ehrenfest relation~\eqref{eq_TE}, to the physical acceleration. We consider the case of a weak acceleration, $a  L_\sigma \ll 1$, so that the physical temperature~\eq{eq_T_x3} is nowhere singular on our lattices. In other words, the position of the Rindler horizon, $z^{\rm R} = - 1/a $, resides far away from the spatial region considered in our simulations, $|z^{\rm R}| \gg L_\sigma$.

We should choose the lattice couplings in Eq.~\eq{eq_S_Karsch} in such a way that the spatial lattice spacing remains constant along the whole lattice,
\begin{align}
    {\mathfrak a}_\sigma = {\mathfrak a}_{0} = \frac{1}{N_\tau T_0} = {\mathrm{const}}\,, 
    \label{a_sigma_fixed} 
\end{align}
while the physical temperature~\eq{eq_T} varies according to Eq.~\eq{eq_T_x3}, implying, in turn, that the temporal lattice spacing is a linear function of the longitudinal coordinate~$z$:
\begin{align}
    {\mathfrak a}_{\tau}(z) = {\mathfrak a}_{0} \bigl(1 + a_0 (z-z_0)\bigr) \,,
    \label{eq_a_tau}
\end{align}
which then corresponds to the anisotropy parameter~\eq{eq_xi}:
\begin{align}
    \xi(z) = \frac{1}{1 + a_0 (z-z_0)}=\frac{T(z)}{T_0}\,.
    \label{eq_xi_x3}
\end{align}

We work at fixed lattice geometry $N_\tau N_\sigma^3$ so that the variation of temperature in the $z$ direction is assumed only in the lattice couplings ${\beta}_\sigma$ and ${\beta}_\tau$ while the lattice extension in the temporal direction, $N_\tau$, is obviously independent of $z$. For the sake of convenience, we take $z_0 = 0$ in Eqs.~\eqref{eq_T_x3}, \eqref{eq_a_tau}, and \eqref{eq_xi_x3}.

A temperature gradient has also been implemented in a recent study of another problem~\cite{Yang:2024tfc} without imposing the fixed-spatial constraint~\eqref{a_sigma_fixed} potentially generating unwanted varying-volume effects. In addition, we notice that the non-linear trigonometric behavior of temperature inhomogeneities introduced in Ref.~\cite{Yang:2024tfc} differs from the thermodynamic formula~\eqref{eq_Ta} given by the solution of the Killing equation~\eqref{eq_Tua}. Therefore, the trigonometric temperature inhomogeneities studied in Ref.~\cite{Yang:2024tfc} do not correspond to the global thermal equilibrium of the accelerating medium studied in our article.

\subsection{Setting up the couplings}

One can consider a linear approximation with Eqs.~(K.2.4), (K.2.24), and (K.2.25) of Ref.~\cite{Karsch:1982ve}, where ``K'' in front of the equation corresponds to this paper by F.~Karsch. It is convenient to parameterize the couplings as follows~\cite{Karsch:1982ve}:
\begin{align}
    {\beta}_\sigma = \frac{6}{g_\sigma^2 ({\mathfrak a},\xi)} \frac{1}{\xi} \,, 
    \qquad
    {\beta}_\tau = \frac{6}{g_\tau^2 ({\mathfrak a},\xi)} \xi \,, 
    \label{eq_beta_sigma_tau}
\end{align}
where at the symmetric value, $\xi = 1$, the couplings are equal:
\begin{align}
    g_\tau ({\mathfrak a},1) = g_\sigma ({\mathfrak a},1) = g_E({\mathfrak a})\,.
\end{align}

For an asymmetric lattice, $\xi \neq 1$, the lattice couplings ${\beta}_\tau$ and $\beta_\sigma$ become different from each other. In the weak coupling limit, relevant to the continuum limit, they can be expanded in terms of the single symmetric coupling~$g_E$:
\begin{subequations}
    \begin{align}
    \frac{1}{g_\sigma^2({\mathfrak a},\xi)} & = \frac{1}{g_E^2({\mathfrak a})} + c_\sigma(\xi) + O(g_E^{3}) \,, \\
    \frac{1}{g_\tau^2({\mathfrak a},\xi)} & = \frac{1}{g_E^2({\mathfrak a})} + c_\tau(\xi) + O(g_E^{3}) \,,
    \end{align}
\label{eq_expansion}
\end{subequations}
~\hskip -5mm where the functions $c_\sigma(\xi)$ and $c_\tau(\xi)$ are given in Ref.~\cite{Karsch:1982ve}. Thus, one should collect all relevant equations of Ref.~\cite{Karsch:1982ve}, neglect $O(g_E^{3})$ terms in Eqs.~\eq{eq_expansion} and use the anisotropy parameter~\eq{eq_xi_x3} to generate the temperature gradient~\eq{eq_T_x3}. However, one should do this carefully. 

We rewrite Eqs.~\eq{eq_beta_sigma_tau} and \eq{eq_expansion} as follows:
\begin{subequations}
    \begin{align}
    {\beta}_\sigma(\beta,\xi) & = \frac{\beta}{\xi} + \frac{6}{\xi} c_\sigma(\xi)\,, \\
    {\beta}_\tau(\beta,\xi) & = \beta \xi + 6 \xi c_\tau(\xi)\,,
    \end{align}
\label{eq_expansion_2}
\end{subequations}
where $O(g_E^{3})$ terms are neglected. We also denoted
\begin{align}
    {\beta}_\sigma = \frac{6}{g_\sigma^2} \frac{1}{\xi}\,, 
    \qquad
    {\beta}_\tau = \frac{6}{g_\tau^2} \xi\,, 
    \qquad 
    \beta = \frac{6}{g_E^2}\,,
\end{align}
where $\beta$ plays a role of the central value of the lattice coupling according to Eqs.~\eq{eq_expansion_2}. From now on, we will work only in terms of $\beta$ and $\xi$.

We must work with the following conditions:
\begin{enumerate}
    \item[(i)] one has anisotropy between spatial and temporal lattice spacings~\eq{eq_xi};
    \item[(ii)] the spatial lattice spacing is fixed regardless of the value of the anisotropy. 
\end{enumerate}
In other words, we have to find the set of the lattice couplings $\beta$ and $\xi$ which fulfill the following conditions:
\begin{enumerate}
    \item[(i)] The anisotropy~\eq{eq_xi} is given by the value of $\xi$:
    \begin{align}
    \xi = \frac{{\mathfrak a}_\sigma(\beta,\xi)}{{\mathfrak a}_\tau(\beta,\xi)}\,.
    \label{eq_condition_i}
\end{align}
    \item[(ii)] At a fixed $\xi$, the value of $\beta$ should be chosen in such a way that the spatial lattice spacing~\eq{a_sigma_fixed} is independent of the asymmetry $\xi$ so that the variations in temperature do not affect the spatial geometry (size of the spatial directions) of the system: 
                \begin{align}
                    {\mathfrak a}_\sigma(\beta,\xi) = {\mathfrak a}_0\,.
                    \label{eq_condition_ii}
                \end{align}
\end{enumerate}
Condition (i) is already ensured by the choice of Eq.~\eq{eq_expansion_2}. However, condition (ii) is less trivial and has to be explicitly cared for.

In the case of the absence of anisotropy, $\xi=1$, Eq.~\eq{eq_condition_ii} means that a fixed value of ${\mathfrak a}$ is achieved at a fixed value of $\beta$. In the presence of a nontrivial anisotropy, $\xi \neq 1$, Eq.~\eq{eq_condition_ii} implies that a fixed ${\mathfrak a}$ constraints a relation $\beta = {\beta}_f(\xi,{\mathfrak a}_0)$ between $\beta$ and $\xi$, such that ${\mathfrak a}_\sigma(\beta(\xi,{\mathfrak a}_0),\xi) = {\mathfrak a}_0$. Here the subscript ``$f$'' in ${\beta}_f$ means ``fixed''. 

Coming back to Ref.~\cite{Karsch:1982ve}, we notice from Eq.~(K.2.6) that the spatial lattice spacing has the following behavior:
\begin{align}
    {\mathfrak a}(\beta,\xi) = {\mathfrak a}(\beta) \exp \left[ - \frac{c_\sigma(\xi) + c_\tau(\xi)}{4 b_0}\right]\,,
    \label{eq_a_beta_xi}
\end{align}
where ${\mathfrak a}(\beta)$ is the lattice spacing for the isotropic lattice, ${\mathfrak a}(\beta) \equiv {\mathfrak a}(\beta,\xi =1)$ and
\begin{align}
    b_0 = \frac{11 N_c}{48 \pi^2} \equiv \frac{11}{16 \pi^2} = 0.69\dots \qquad \text{for} \quad N_c = 3\,,
\end{align}
is the first coefficient of the beta function. According to Ref.~\cite{Karsch:1982ve}, the correction in the spatial lattice spacing~\eq{eq_a_beta_xi} due to non-trivial behavior of $\xi$ is very small, of the order of 1\%. The physical values of the isotropic lattice coupling ${\mathfrak a}(\beta)$ in Eq.~\eq{eq_a_beta_xi} as functions of $\beta$ can be found in Table~1 of Ref.~\cite{Athenodorou:2020ani}. 

The final functions defining the $\beta(\xi)$ dependence are presented below:
\begin{subequations}
    \begin{align}
    \label{eq_beta_inuse_a}
    {\beta}_\sigma({\beta}_0,\xi) & = \frac{1}{\xi}\lr{\beta({\beta}_0,\xi) + 6\cdot 1.105\cdot c_\sigma(\xi)}\,, \\
    \label{eq_beta_inuse_b}
    {\beta}_\tau({\beta}_0,\xi) & = \xi\lr{\beta({\beta}_0,\xi) + 6\cdot c_\tau(\xi)}\,, \\
    \label{eq_beta_inuse_c}
    \beta({\beta}_0,\xi) & = \beta\lr{{\mathfrak a}({\beta}_0) \exp \left[\frac{c_\sigma(\xi) + c_\tau(\xi)}{4 b_0}\right]}\,,
    \end{align}
    \label{eq_beta_inuse}
\end{subequations}
where function $\beta\lr{{\mathfrak a}}$ is the inverse function for ${\mathfrak a}\lr{\beta}$~\cite{Athenodorou:2020ani}, ${\beta}_0$ is the critical $\beta$-value for finite temperature. We performed calculations with the lattice size in temporal direction $N_t=8$ (${\beta}_0=6.07$).
In function~(\ref{eq_beta_inuse_a}) we add the additional multiplication factor $1.105$ for improve stability of the ${\mathfrak a}_\sigma={\mathrm{const}}$ in the used $\xi$-range. We got the value of the factor by fitting lattice data.

\begin{figure}[!h]
  \begin{center}
    \includegraphics[width=1.0\linewidth]{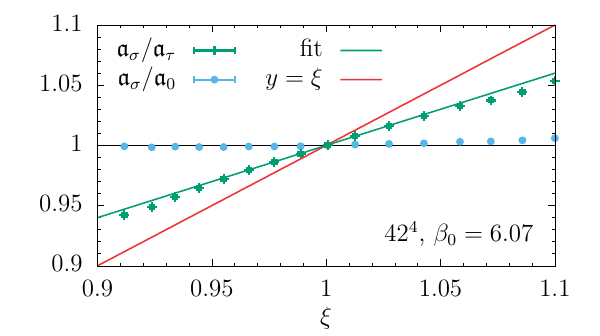}
  \end{center}
  \vspace{-6ex}
  \caption{Relations ${\mathfrak a}_\sigma/{\mathfrak a}_\tau$ and ${\mathfrak a}_\sigma/{\mathfrak a}_0$ are obtained from the analysis of the Creutz ratio for Wilson loops $5\times 5$ for ${\beta}_0=6.07$ on the lattice $42^4$. We used the line function for fit: $y(\xi)=k\xi+(1-k)$ which gives us the following best fit value: $k=0.600(16)$.}
  \label{fig_CreutzRatio}
\end{figure}

At zero temperature, we explicitly calculated the asymmetry value $\xi$ from the Creutz ratio shown in Fig.~\ref{fig_CreutzRatio}. We kept the spatial size of the lattice constant, and the resulting deviations were, again, less than one percent.

\subsection{\add{Infinite-volume limit and the nature of the transition}}

\begin{figure}[!h]
\begin{tabular}{c c}
  \includegraphics[width=.42\linewidth]{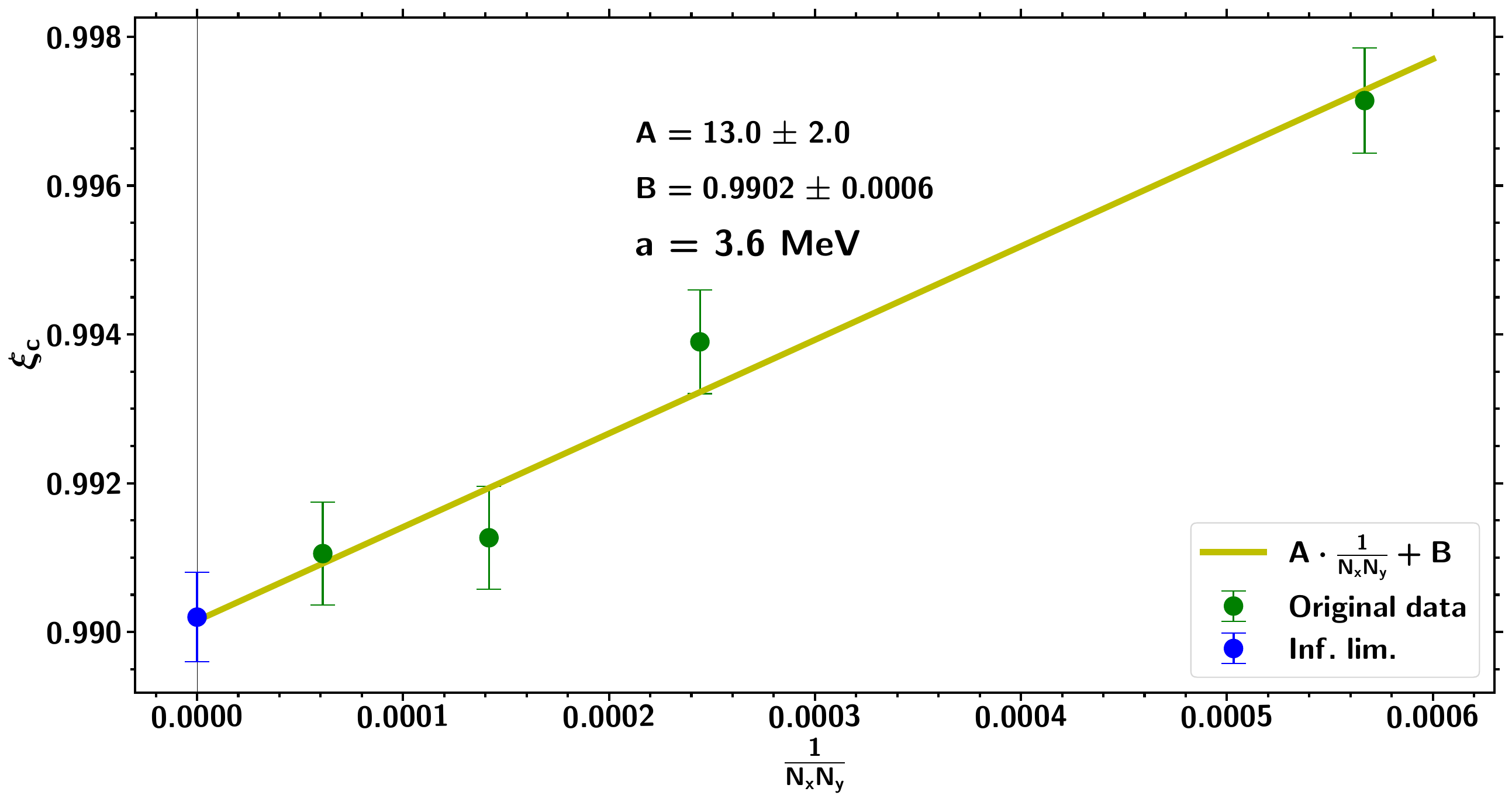} & $\qquad$
  \includegraphics[width=.42\linewidth]{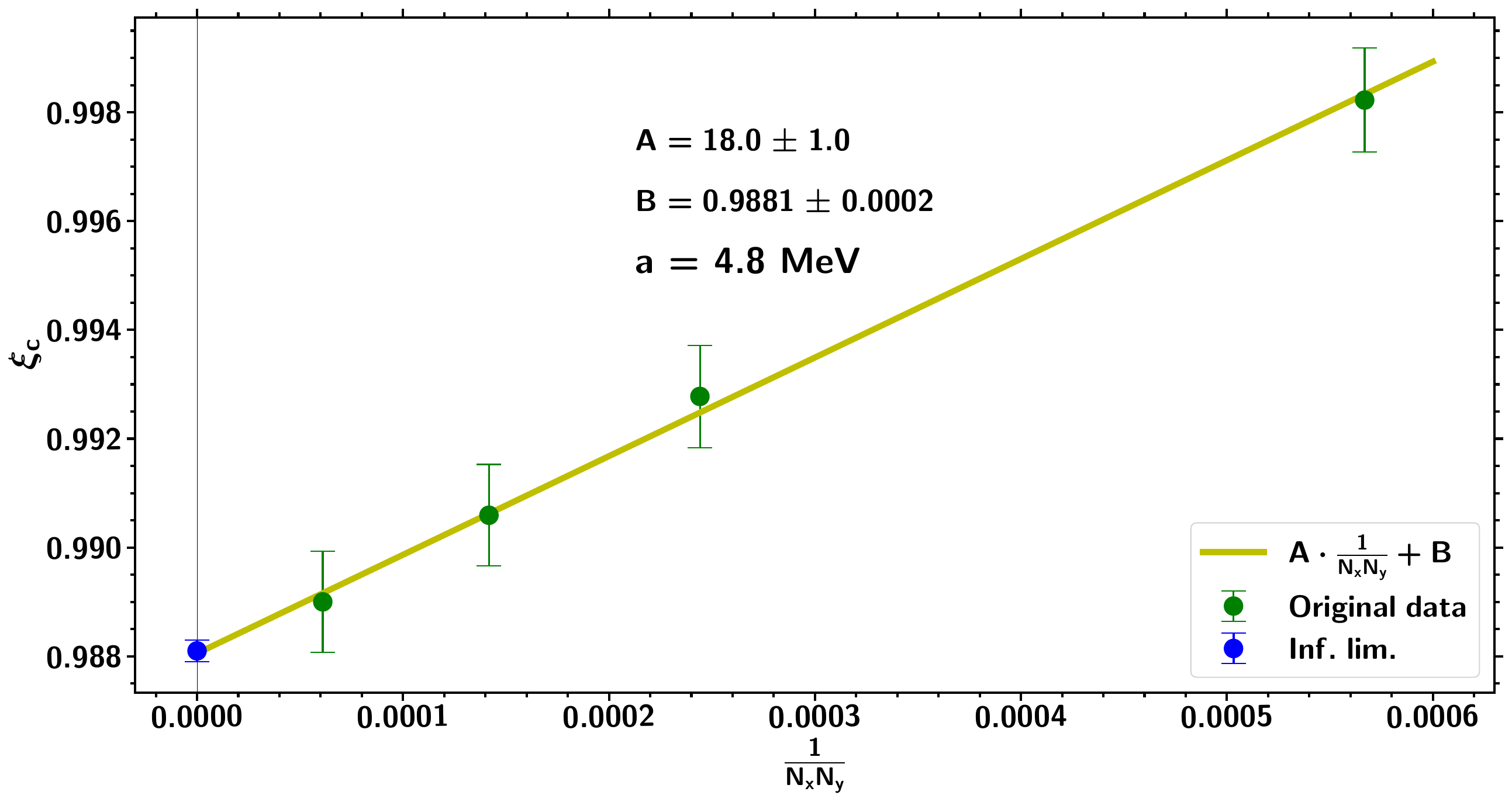} \\
  \includegraphics[width=.42\linewidth]{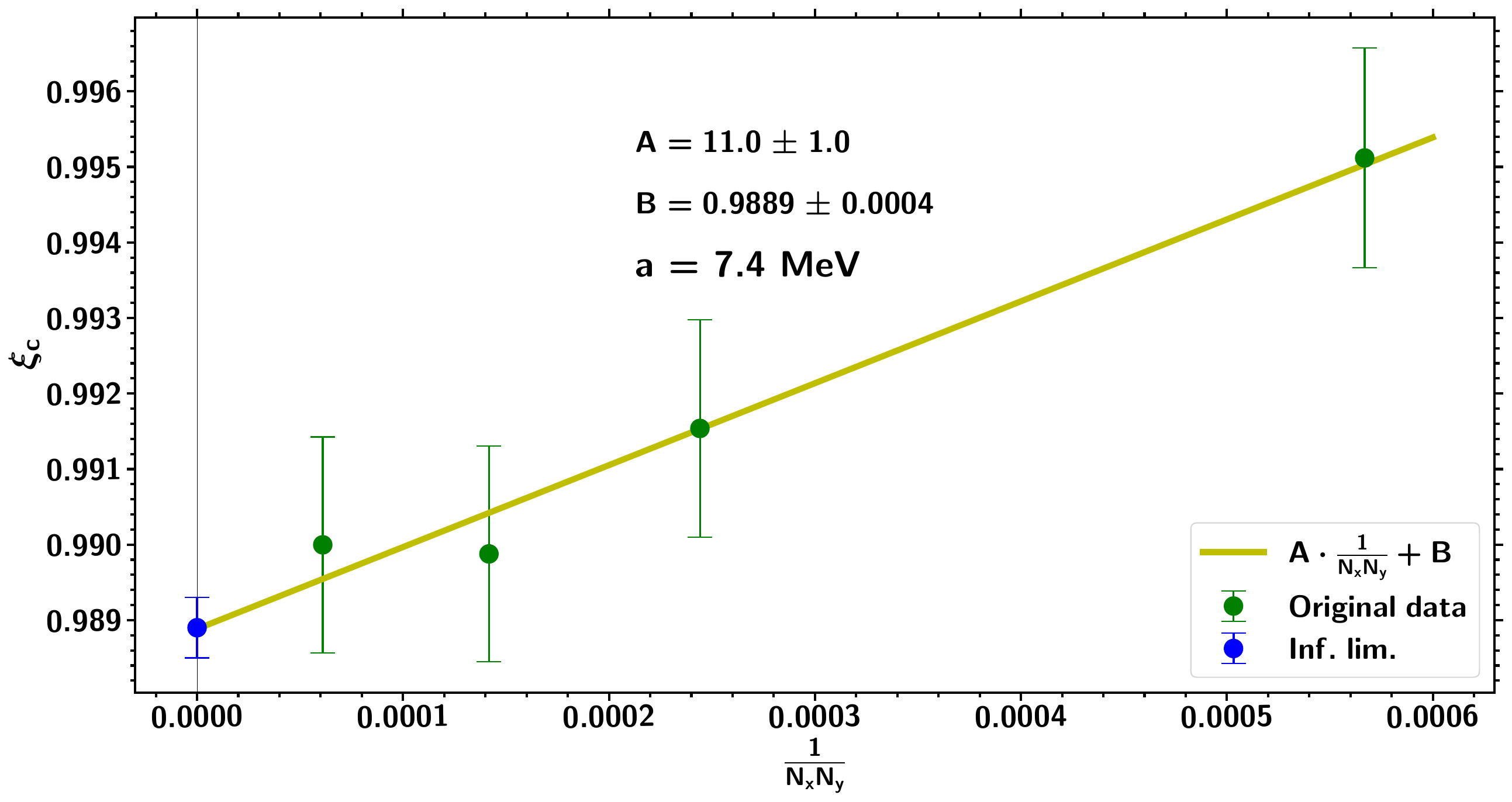} &  $\qquad$
  \includegraphics[width=.42\linewidth]{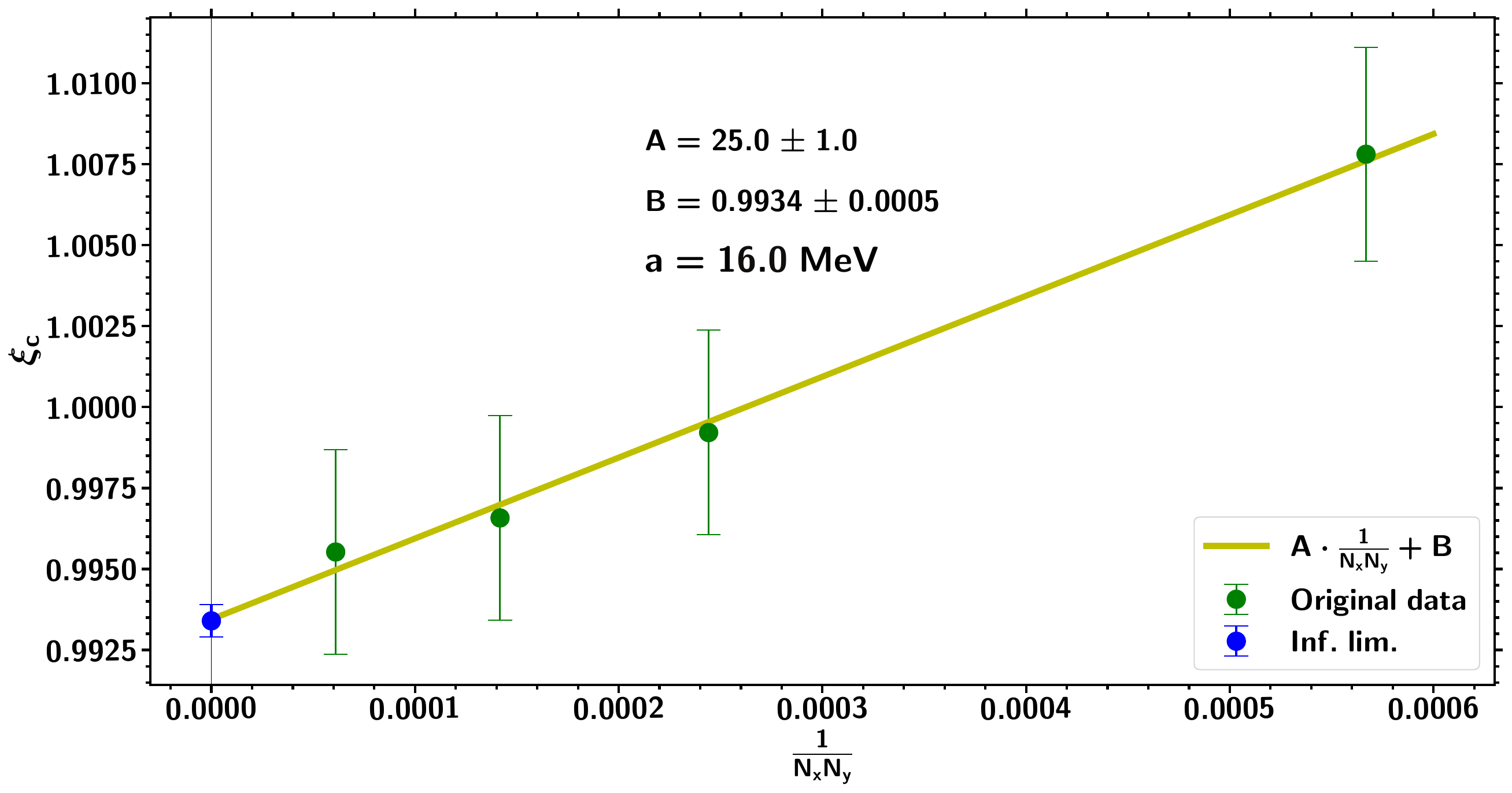}
\end{tabular}

  \vspace*{-2ex}\centering
  \caption{\add{Infinite-volume limit in the $xy$-plane for the peak position of the susceptibility of the Polyakov loop in terms of the critical anisotropy parameter $\xi_c$. The data, shown by the green points, correspond to the transverse extensions $N_x=N_y=42,64,84$, and $128$. The linear fitting functions $\xi_c (N_x, N_y) = A/(N_x N_y) + B$ and the best fit parameters $A$ and $B$ are shown in the insets of the plots. The infinite-volume limit is shown by the blue point at $1/(N_x N_y) \to 0$.}}
  \label{fig_InfLimXYplane}
\end{figure}

\add{The positions of the crossover peaks of the Polyakov loop susceptibility shown in Fig.~\ref{fig_Polyakov_susceptibility} are translated into the critical values of the anisotropy parameter $\xi = \xi_a$ for each lattice volume. The infinite-volume limit is obtained by sending the transverse size of the lattice to infinity, $N_x = N_y \to \infty$. The positions of the peaks in the anisotropy parameter $\xi$ scale linearly with the decreasing inverse lattice volume, as the data can be well described by the linear function $\xi_c (N_x, N_y) = A/(N_x N_y) + B$ at any studied acceleration. Here $A$ and $B$ are the fitting parameters, where the parameter $B$ represents the infinite-volume value of the quantity $\xi_c$. This linear behavior exhibits a well-defined infinite-volume limit with finite values of $B$, shown, together with the linear fits, in Fig.~\ref{fig_InfLimXYplane}.}

\begin{figure}[!h]
\begin{tabular}{cc}
  \includegraphics[width=.42\linewidth]{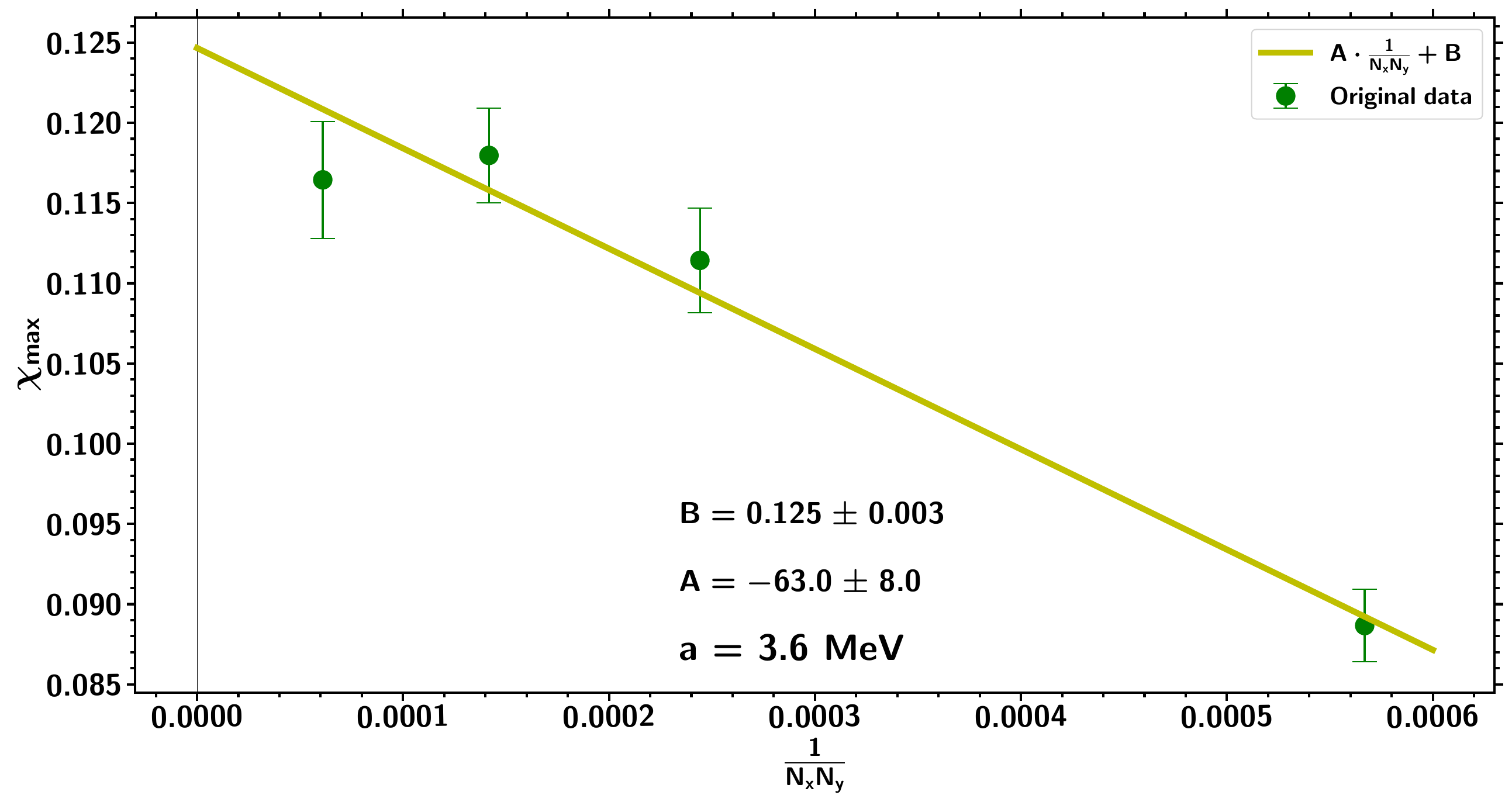} & $\qquad$
  \includegraphics[width=.42\linewidth]{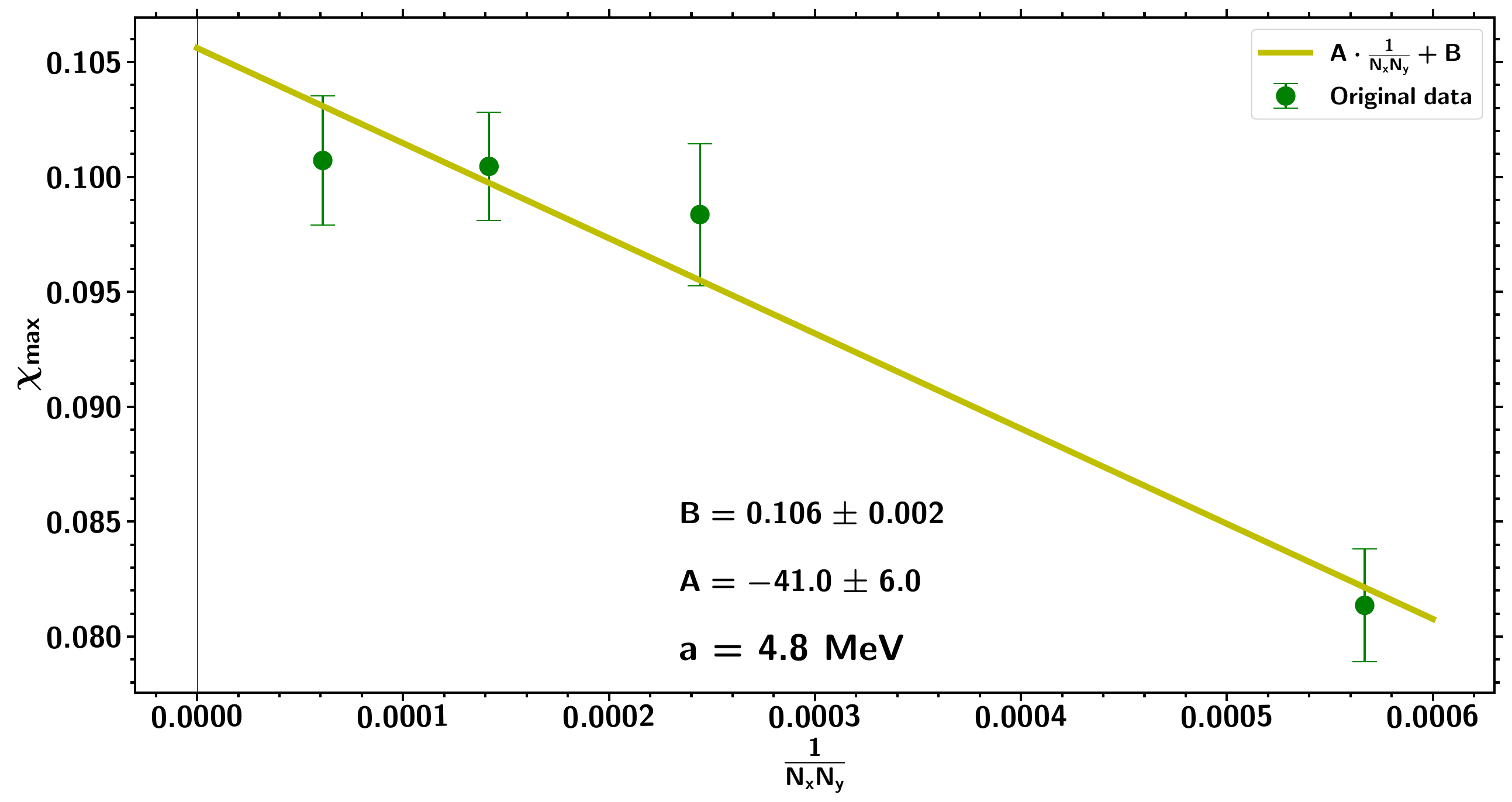} \\
  \includegraphics[width=.42\linewidth]{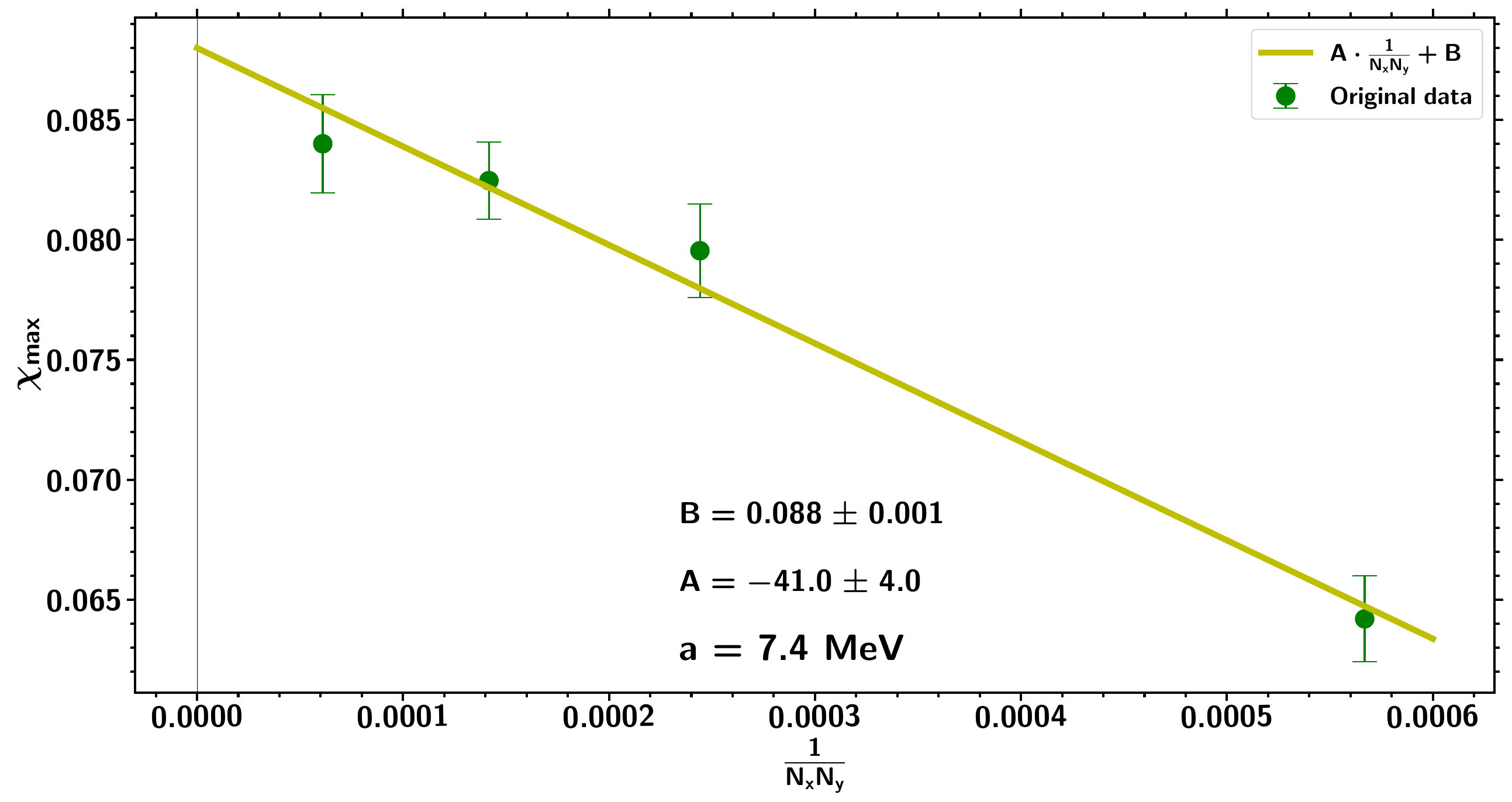} &  $\qquad$
  \includegraphics[width=.42\linewidth]{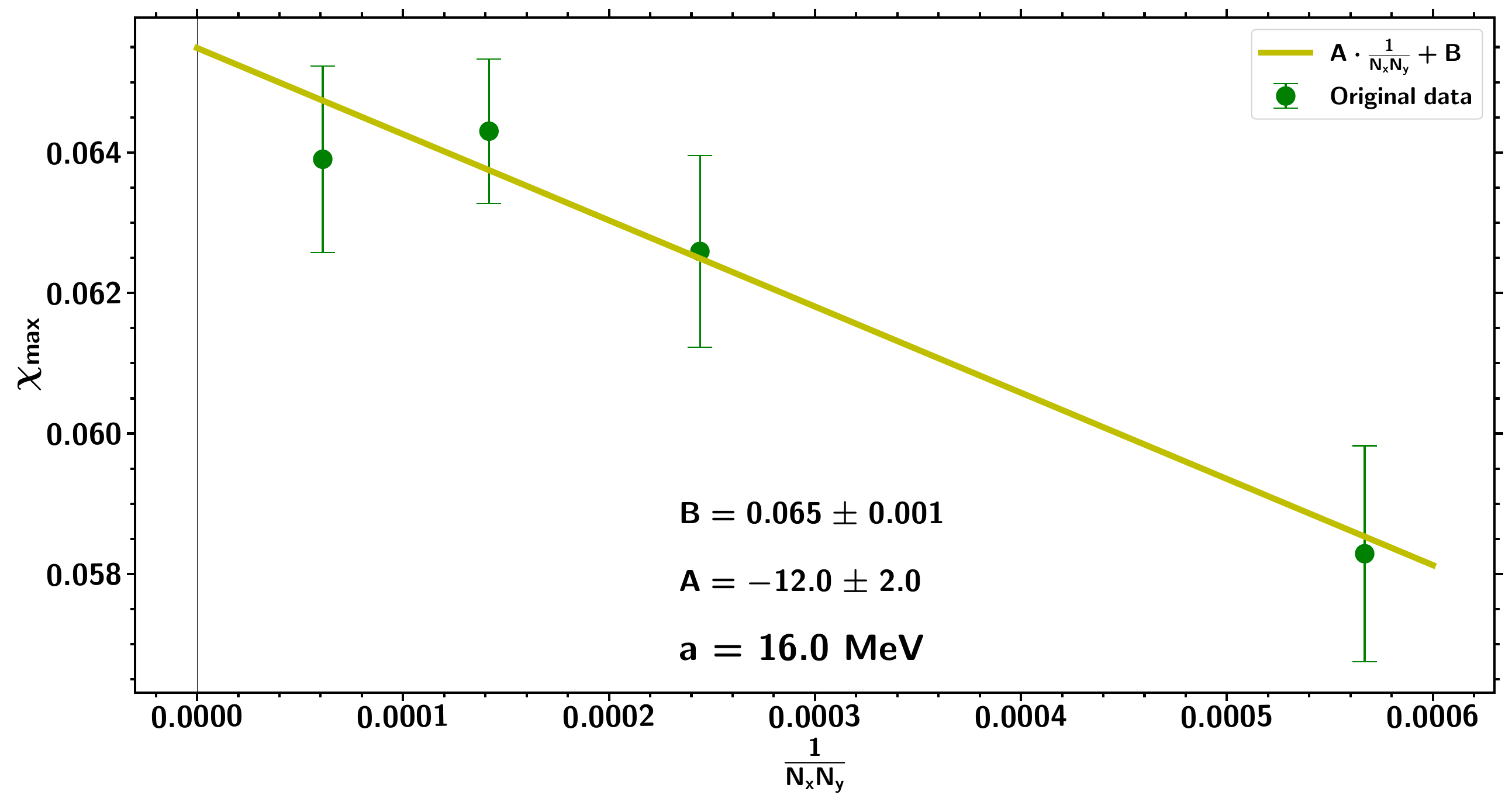}
\end{tabular}

  \centering
  \caption{\add{The same in Fig.~\ref{fig_InfLimXYplane} but for the height of the peak of the susceptibility of the Polyakov loop.}}
  \label{fig_maxXI}
\end{figure}

\add{The nature of the transition in the accelerating system can be revealed with the help of the scaling of the height of the peak $\chi_{\rm max}$ in the Polyakov loop susceptibility as a function of the spatial volume. In Fig.~\ref{fig_maxXI}, we show this quantity  for various accelerations. Interestingly, the data for the height can also be described by a linear function of the inverse volume, $\chi_{\max} (N_x, N_y) = A/(N_x N_y) + B$ where $A$ and $B$ are other fitting parameters. The linear behavior implies that the height of the peak has a non-singular behavior in the infinite-volume limit, thus indicating the absence of a true phase transition for all studied values of acceleration.}

\begin{figure}[!h]
\begin{tabular}{cc}
  \includegraphics[width=.42\linewidth]{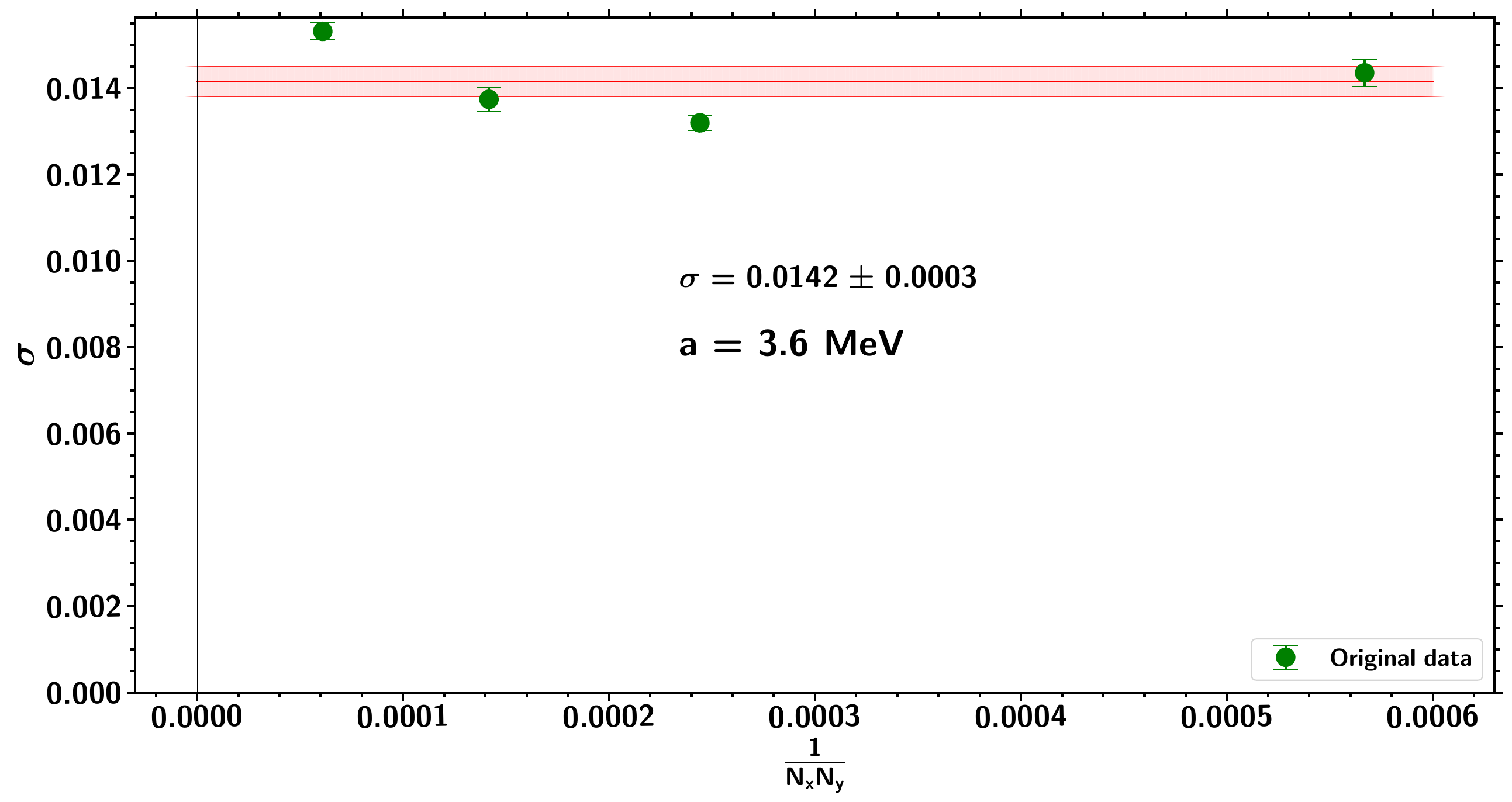} & $\qquad$
  \includegraphics[width=.42\linewidth]{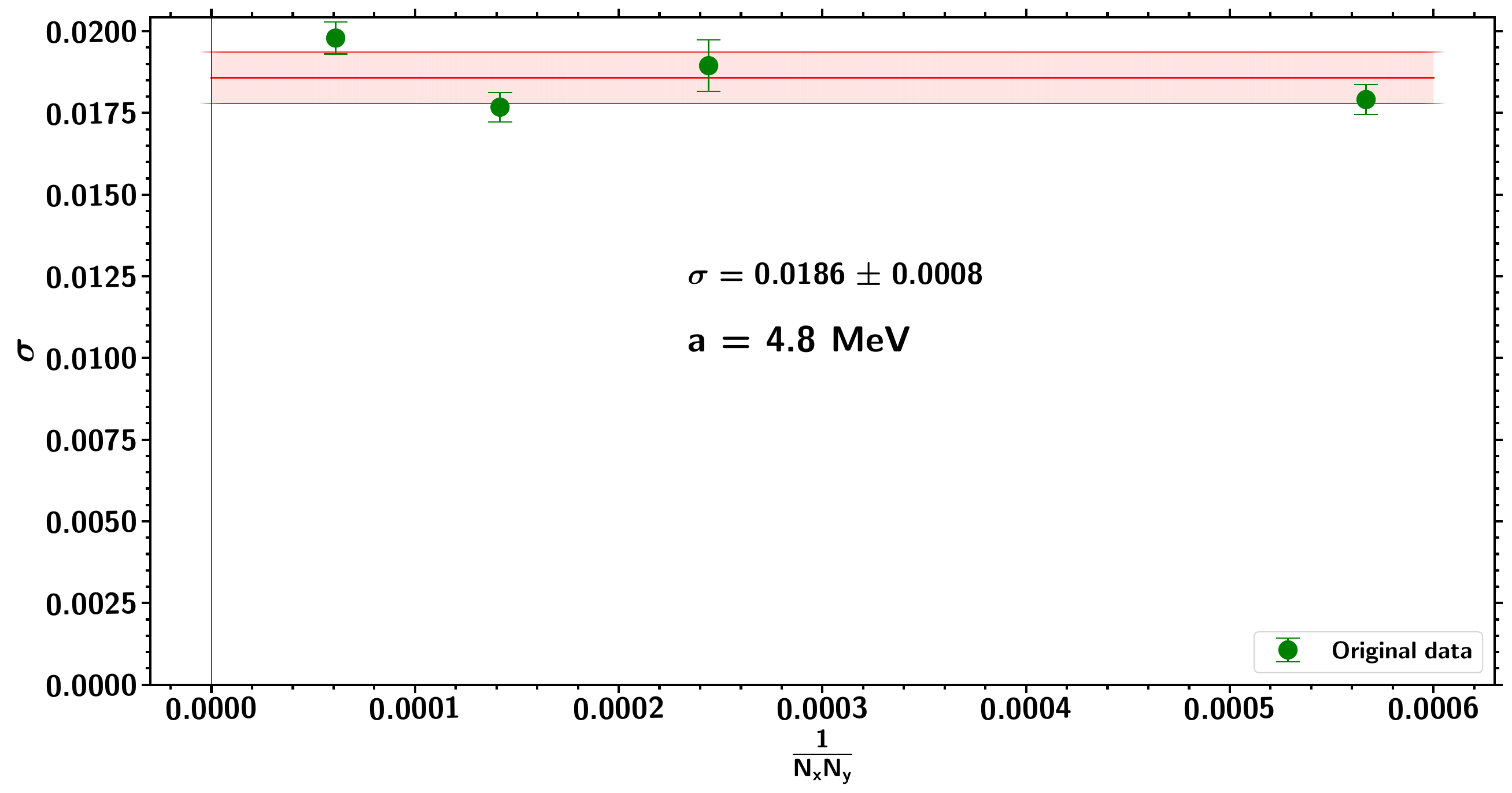} \\
  \includegraphics[width=.42\linewidth]{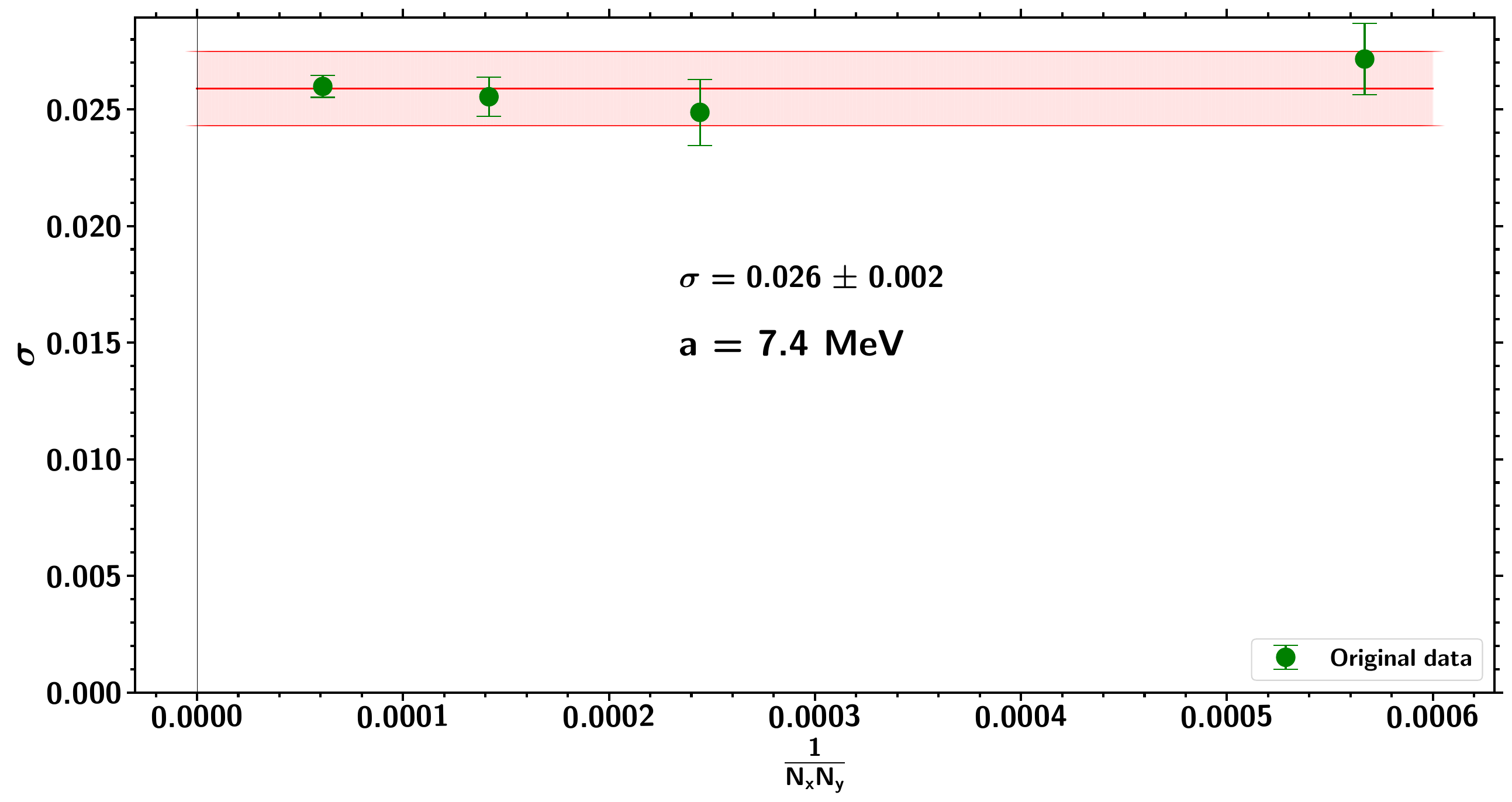} &  $\qquad$
  \includegraphics[width=.42\linewidth]{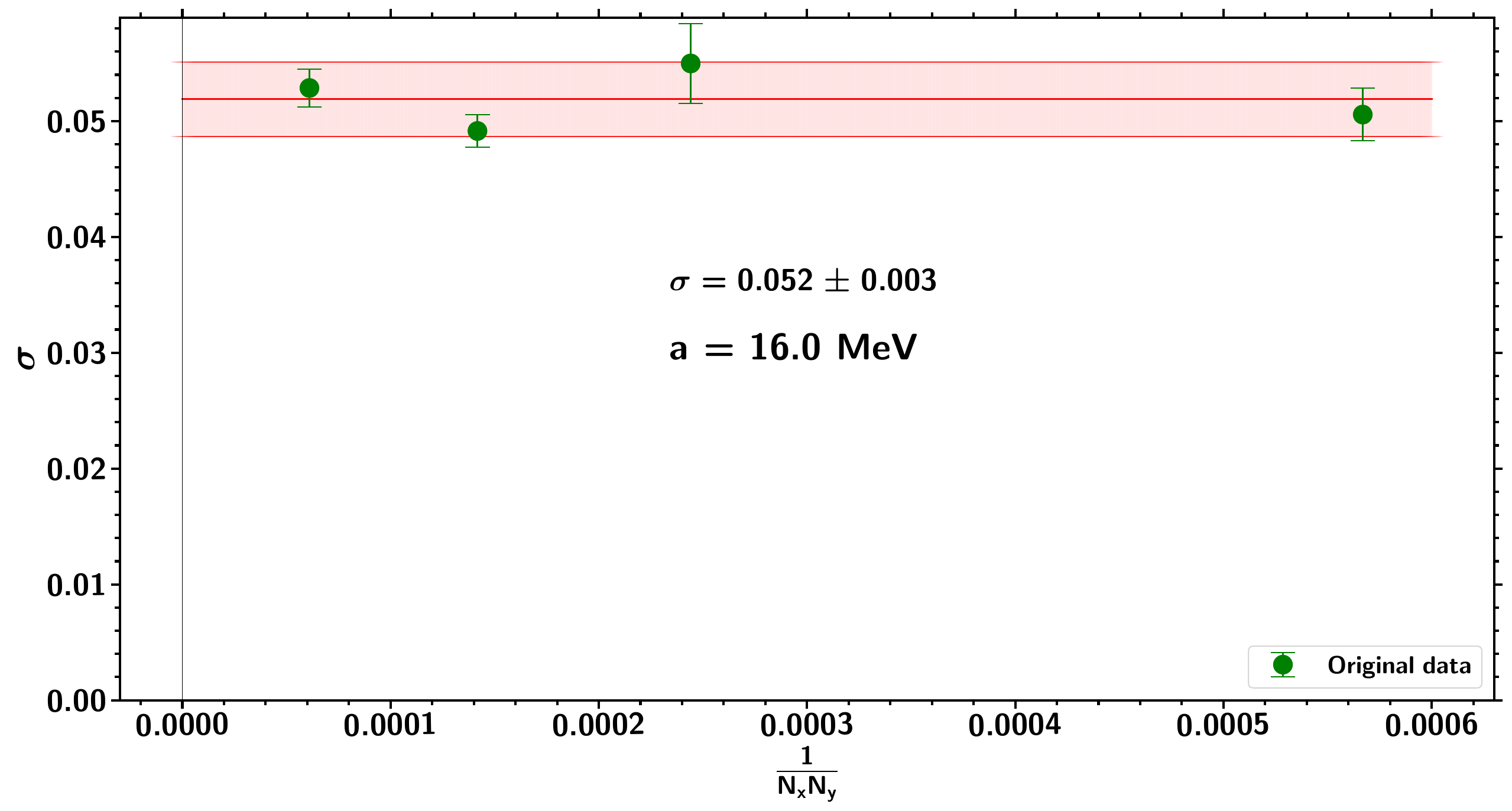}
\end{tabular}

  \centering
  \caption{\add{The width of the Polyakov loop susceptibility $\sigma$ as the function of the inverse transverse area $1/(N_x N_y)$ for various accelerations $a$. The red horizontal line denotes the average of the susceptibility width $\bar\sigma$, with the shaded region denoting a statistical error $\delta \sigma$ in the width $\sigma = \bar\sigma \pm \delta \sigma$ (in each plot, these quantities are shown explicitly).}}
  \label{fig_Sigma}
\end{figure}

\add{For a true thermodynamic phase transition, the height of the susceptibility peak should rise with the volume while the peak itself should become narrower. In other words, the width of the susceptibility peak $\sigma$ is expected to get smaller as the volume rises. For a crossover, the width is expected to be roughly independent of the volume. We checked this assertion in Fig.~\ref{fig_Sigma}, where the susceptibility width $\sigma$ is shown as the function of the inverse transverse area $1/(N_x N_y)$ of the lattice with all other lattice dimensions kept fixed. The observed independence of $\sigma$ on the volume of the system is consistent with the earlier observation of the non-singular scaling of the height $\chi_{\rm max}$ of the susceptibility peak. Both these numerical results indicate that the transition for all studied non-vanishing accelerations is a smooth crossover.}

\onecolumngrid

\vspace*{2ex}

\vspace*{2ex}

\vspace*{6ex}


\end{document}